\documentclass[final]{agujournal2019}
\usepackage{url} 
\usepackage{lineno}
\usepackage{floatrow,bm}
\usepackage{soul}
\usepackage{amsmath,amssymb}
\usepackage{accents}
\newcommand{\ubar}[1]{\underaccent{\bar}{#1}}
\newcommand{\uubar}[1]{\ubar{\ubar{#1}}}

\usepackage{enumitem}
\journalname{}

\begin{document}

%
%


\title{Control instabilities and incite slow-slip in generalized Burridge-Knopoff models}

%
%




\authors{Ioannis Stefanou}
\affiliation{1}{
 Ecole Centrale de Nantes, Universit\'e de Nantes, CNRS \\ GeM (Institut de Recherche en G\'enie Civil et M\'ecanique), Nantes, France
}%





\correspondingauthor{Ioannis Stefanou}{ioannis.stefanou@ec-nantes.fr}

\begin{abstract}
Generalized Burridge-Knopoff (GBK) models display rich dynamics, characterized by instabilities and multiple bifurcations. GBK models consist of interconnected masses that can slide on a rough surface under friction. All masses are connected to a plate, which slowly provides energy to the system. The system displays long periods of quiescence, interrupted by fast, dynamic events (avalanches) of energy relaxation. During these events, clusters of blocks slide abruptly, simulating seismic slip and earthquake rupture.\\
Here we propose a theory for preventing GBK avalanches, control its dynamics and incite slow-slip. We exploit the dependence of friction on pressure and use it as a backdoor for altering the dynamics of the system. We use the mathematical Theory of Control and, for the first time, we succeed in (a) stabilizing and restricting chaos in GBK models, (b) guaranteeing slow frictional dissipation and (c) tuning the GBK system toward desirable global asymptotic equilibria of lower energy. Our control approach is robust and does not require exact knowledge of the frictional behavior of the system. Finally, GBK models are known to present Self-Organized Critical (SOC) behavior. Therefore, the presented methodology shows an additional example of SOC Control (SOCC).\\
Given that the dynamics of GBK models show many analogies with earthquakes, we expect to inspire earthquake mitigation strategies regarding anthropogenic and/or natural seismicity. In a wider perspective, our control approach could be used for improving understanding of cascade failures in complex systems in geophysics, access hidden characteristics and improve their predictability by controlling their spatio-temporal behavior in real-time.
\end{abstract}

\section{Introduction}
\label{sec:intro}
Generalized Burridge-Knopoff models (GBK) \cite{Burridge1967} models consist of several interconnected masses that can slide on a rough surface under friction. The sliding blocks are connected to a driver plate that slowly transfers load and energy to the system. This system is characterized by slow and fast dynamics and shows spatio-temporal correlations due to the high interconnectivity of the masses. The continuous slow movement of the driver plate leads to instabilities expressed as abrupt sliding of clusters of blocks, which can be avoided following the proposed theory. GBK models are also frequently used as simplified analog models for earthquakes, as they combine the basic mechanisms of Reid's elastic rebound theory \cite{Reid} and statistical similarities.

The main objective of this work is to show that the chaotic, rich dynamics of generalized Burridge-Knopoff models can be altered and controlled. Then, slow-slip and smooth energy relaxation can be achieved in a controlled and designed manner. Our theory is inspired by the recent experience that humans \textit{do cause} earthquakes by injecting fluids under pressure in the earth's crust leading to fault reactivation \cite{Raleigh1976,Mcgarr2002,Keranen2013a,Rubinstein2015,Cornet2016,
Guglielmi2015a,Cappa2019}. As a result of fluid pressure change, the spatio-temporal distribution of earthquakes can be perturbed \cite<e.g.>{Petersen2015a}. However, until now, this is done in an uncontrollable manner, which results in significant criticism of active projects involving fluid injections in the earth's crust (see conventional and unconventional energy production in oil and gas industry, renewable energies like deep geothermal projects, $\textrm{CO}_2$ sequestration etc.). Here, the problem of fluid injections is seen from another perspective. 

We exploit the frictional dependence on pressure and the possibility of altering the local equilibrium conditions by fluid injections. Consequently, fluid pressure is considered as a backdoor for altering the dynamics of the GBK system. 
The main ingredient of our approach is the \textit{modern mathematical Theory of Control} \cite{Vardulakis1991,Vardulakis2012,Ka}, which is applied in order to:
\begin{enumerate}[noitemsep, topsep=0pt, partopsep=0pt]
  \item Stabilize the GBK system and
  \item drive it to lower energy levels and stable equilibria.
\end{enumerate} 
This is achieved without precise knowledge of the mechanical parameters of the system and of its heterogeneous and uncertain frictional properties.

GBK models are considered as qualitative analogs of a single earthquake fault or as models of distributed seismicity \cite<see>[among others]{Turcotte1999}.
However, it is worth emphasizing that the dynamic behavior of real faults can be much richer than that of the Burridge-Knopoff model and of its generalization \cite<cf.>[]{Barbot2019, Barbot2012}. The main limitations of models of the Burridge-Knopoff type are extensively discussed and shown by \citeA{Rice1993} and later publications. However, the value of using this analog model resorts to its simplicity and the fact that its rich dynamic behavior is well documented and thoroughly discussed in the literature. Moreover, its mathematical structure allows the development of a general control approach that could be applied in more realistic situations of earthquake rupture and instability, provided that a consistent discretization approach like the one proposed by \citeA{Rice1993,Chinnery1963,Erickson2011} is followed (see also \ref{sec:equations_of_motion}). Therefore, from the mathematical point of view, more realistic cases could be tackled using the theoretical developments presented herein, but this extends the scope of the present work. However, it is worth mentioning that the present approach has several limitations. These are related, among others, to the actual techno-economical means for fluid injections in the earth's crust, the sampling-rate and frequency of observations, the exclusion of poroelastic effects and the in-situ hydrogeological and geomechanical conditions. These limitations and their solution is explored in the frame of the ongoing ERC project ``Controlling earthQuakes - CoQuake'' (\url{http://coquake.eu}).

The paper is organized as follows. In Section 2 we present the main ingredients of generalized Burridge-Knopoff models, we discuss their frictional properties and we show how the dependence of friction on pressure can be exploited for achieving robust control of the dynamics of the system. In Section 3 we confirm the well-known \textit{Self-organized critical} (SOC) behavior for GBK models and we give an example of system stabilization using our theoretical developments. Furthermore, we show how the input pressure can be adjusted in real-time for assuring slow-slip and driving the system to lower (potential) energy states (robust tracking). The mathematical proofs for stabilization and tracking under uncertainties are given in the Appendices. The results of our approach are discussed extensively in the last Section of this work, where perspectives and implications of the current framework are given for SOC control and geophysics, in general, and for man-made and natural earthquakes, in particular.

\section{Theoretical model}
\label{sec:theoretical_model}
\subsection{Generalized Burridge-Knopoff model's dynamics}
\label{sec:GBK_dynamics}
\begin{figure*}
\centering
\includegraphics[width=1.0\linewidth]{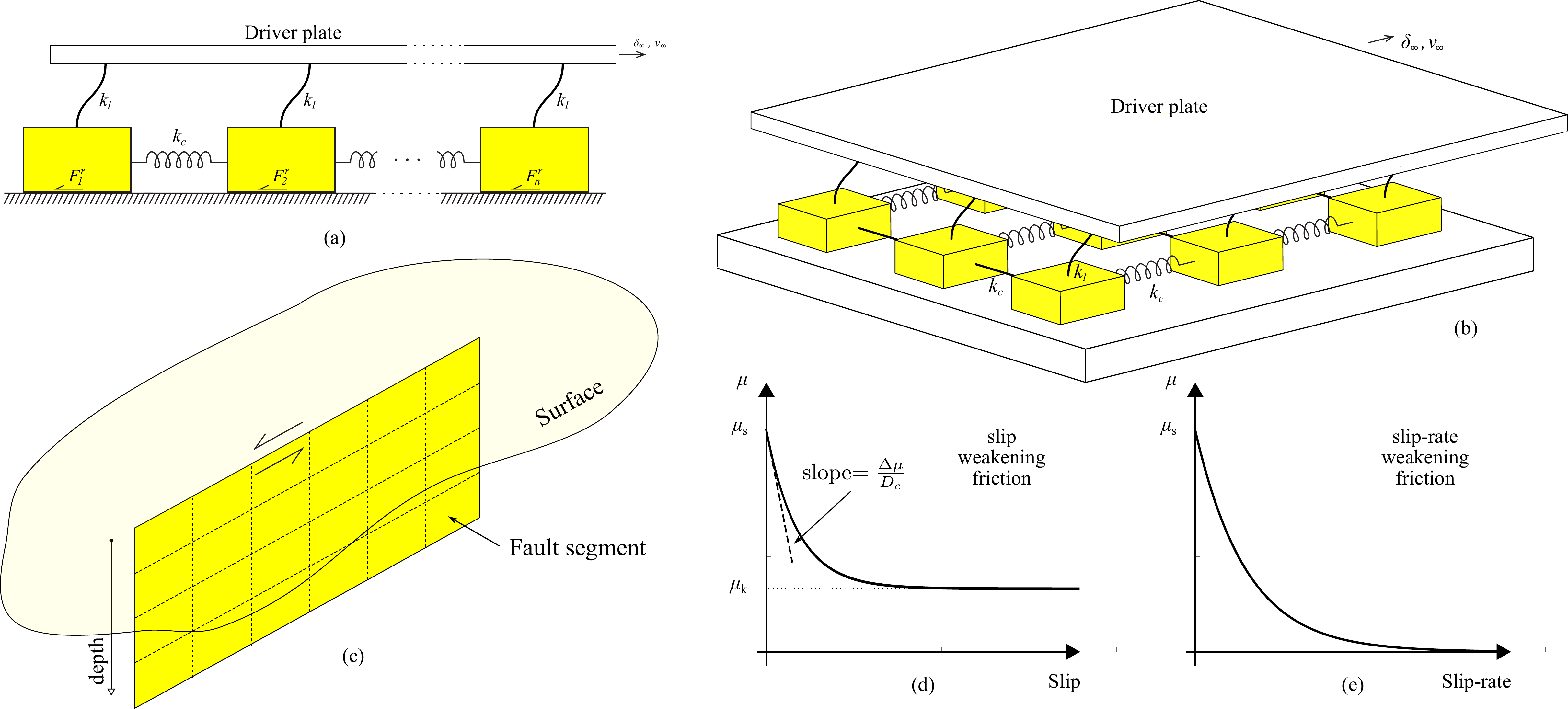}
\caption{Schematic representation of (a) the Burridge-Knopoff model (b) its two-dimensional generalization and (c) of a strike-slip fault discretized in $N_x \times N_z$ segments ($N_x=4$, $N_z=6$ in this Figure). Evolution of friction coefficient with slip (d) and slip-rate (e). The dashpots considered in the mathematical model of the current GBK model are not drawn for the sake of simplicity.}
\label{fig:gen_slider}
\end{figure*}

We consider an ensemble of $n$ blocks of mass $m$ each one, connected through springs and dampers. Each block can slide independently on a rough, horizontal plane as shown in Fig.~\ref{fig:gen_slider}. The blocks are connected together and with a plate through springs and dampers. The plate, which is called here \textit{driver plate}, is translated under constant velocity and provides energy to the system, which is progressively stored in the elastic springs. During this phase the system is in stable equilibrium and its total potential energy increases. Due to slip and slip-rate frictional weakening (Fig.~\ref{fig:gen_slider}d,e), at a certain point this equilibrium becomes unstable and some blocks slide abruptly. During this phase, a part of the stored energy is dissipated abruptly due to friction and damping. The dynamics of this system is described by the following set of non-dimensional equations (see \ref{sec:equations_of_motion}):
\begin{equation}
\ubar{x}'= \uubar{G}\ubar{x}+\ubar{H}(\ubar{x}),
\label{eqn:dynamics_simple}
\end{equation}
where $\ubar{x}$ is the state vector, whose first $n$ components represent the dimensionless displacements and the rest $n$ components the dimensionless velocities of the masses, $(.)'$ denotes the dimensionless time derivative, the vector $\ubar{H}(\ubar{x})$ represents the forces applied to the blocks due to the initial deformation of the springs, the displacement of the driver plate and friction. The matrix $\uubar{G}$ contains information on how the blocks are connected together and can describe different physical situations as shown in Fig.~\ref{fig:gen_slider}. These physical situations span from the classical Burridge-Knopoff model, \cite<see Fig.~\ref{fig:gen_slider}a and>[among others]{Burridge1967,Dieterich1972,Carlson1989}, its 2D generalization \cite<see Fig.~\ref{fig:gen_slider}b and>[among others]{Ito1990,Brown1991,Huang1992} and a strike-slip fault discretized in $N_x \times N_z$ segments (Fig.~\ref{fig:gen_slider}c). The system is said to be in equilibrium when $\ubar{x}'=0$. All the physical quantities of the system were scaled as described in \ref{sec:equations_of_motion}, where more details are also given. 

It is worth emphasizing that several variants of the original Burridge-Knopoff model exist in the literature \cite<>[among many others]{Ben-zion2008,DeArcangelis2016}. For this purpose, the above formulation is kept general and allows the incorporation of most of these variants by adapting the connectivity matrix $\uubar{G}$ and the forcing vector $\ubar{H}$. This general formulation allows to derive general proofs about the control of the system as shown in \ref{sec:control_stabilization} and \ref{sec:control_tracking}. These proofs account for robustness, meaning that the system can be controlled even in the absence of detailed information regarding the connectivity of the blocks (including elasticity and viscosity) and friction. Here we considered only uncertainties related to friction, but the extension of the theory to cover uncertainties related to elasticity, viscosity and connectivity is straightforward. 

\subsection{Friction and instabilities}
We assume Coulomb friction, $F_i^r = \mu_i N_i^{(e)}$ , where $N^{(e)}_{i}$ is the effective normal force applied on block $i$ and $\mu_i$ is the friction coefficient that may depend on slip, slip-rate, time and other internal/state variables. 

In the frame of fault mechanics, several rheological models have been proposed for describing the apparent friction of faults and fault gouges \cite<e.g.> {Byerlee1978,Dieterich1981,Scholz2002}. In tribology as well. Despite the existing wide experimental and theoretical knowledge about friction and given the presence of heterogeneities \cite<cf.>{Dieterich1979,Marone1998,DiToro2011, Rattez2018a,Rattez2018b,Barras2019,Rattez2020,Kenigsberg2020,Collins-craft2020}, friction is not a well constrained quantity and it is hard to quantify without large uncertainties. To this extent we keep here a simple, but general rheology for friction and we adopt a static/dynamic friction law. In particular, the coefficient of friction evolves from an initial value $\mu_s$ (static friction coefficient), to a residual one $\mu_k$ (kinetic friction coefficient). Fig.~\ref{fig:gen_slider}d-c shows schematically the transition between static and kinetic friction for a slip and slip-rate weakening law. This transition is made in a characteristic distance $D_c$ \cite{Kanamori2004}. This characteristic distance, together with the friction coefficient drop and the applied effective normal force, determine the apparent frictional weakening during sliding and render the system unstable leading to sudden slip \cite<stick-slip motion>{Ruina1983, Stefanou2019}. Alternatively, the classical rate-and-state friction law could be used, but this model is mathematically singular for very low (and very high) sliding velocities, as the ones developed herein when the system is under control, and therefore it is not appropriate without proper regularization \cite{Ben-zion2008}.

From the physical point of view, a dynamic instability takes place when the elastic unloading of the springs cannot be counterbalanced by friction. The same holds when the friction shows velocity weakening that cannot be counterbalanced by the viscosity of the dashpots. Stability analysis using Lyapunov methods can show the exact conditions for which this system becomes unstable \cite<cf.>[]{Stefanou2019}. When several blocks are considered, the system presents a chaotic, SOC behavior, which makes its dynamical behavior extremely rich and challenging to control \cite<see for instance>[]{Carlson1989,Huang1992,Schmittbuhl1996,Becker2000,Erickson2011}.

\subsection{Input and robust control}
\label{sec:friction}
Key element for the control and arrest of the above mentioned instability is the effective normal force $N_i^{(e)}$, which can give us a valuable access to the dynamics of the system (backdoor). By changing $N_i^{(e)}$, the apparent friction of the blocks can be modified in a desired way. For instance, by reducing $N_i^{(e)}$, the friction is reduced resulting in a lubrication-like effect and enhancing slip. By increasing $N_i^{(e)}$, the apparent frictional force is increased and hampers sliding. These mechanisms are central for stabilizing the system and driving it smoothly to a desired equilibrium of lower energy, using the mathematical theory of control as shown below.

For instance, the effective normal force and, therefore, the apparent friction, can be altered by injecting fluids into the frictional interfaces where sliding takes place. The injection of fluids can change the fluid pressure and, therefore, the apparent friction. Adopting Terzaghi's principle of effective stress \cite{Terzaghi1925}, $N_i^{(e)}=N_i^0-P_i$, where $P_i$ is the interstitial fluid pressure change at the interface of block $i$ and $N_i^0$ is a constant, reference normal force (e.g. the weight of the block). As a result, fluid pressure can play the role of \textit{input} into the dynamical system, described by Eq. (\ref{eqn:dynamics_simple}).

In the case of real faults, $N_i^0$ is a fraction of the effective overburden earth load, averaged over the area of the faut's segment $i$. Its value depends on the tectonic setting and the type of the fault. $P_i$ corresponds to a change in the interstitial fluid pressure, provoked, for example, by fluid injections at the vicinity of the segment $i$. It is worth mentioning that injecting fluids in the earth's crust and altering the local equilibrium by fluid pressure changes is nowadays a common practice in many industrial projects. Some examples are deep geothermal projects, $\textrm{CO}_2$ sequestration and the oil industry \cite{Rubinstein2015}. However, recent experience shows \cite<e.g.>{Keranen2013a,Cornet2016,Grigoli2018, Kwiatek2019,Cornet2019,Parisio2019,Hofmann2019,Cauchie2020} that, in many cases, earthquakes are triggered due to the reactivation of tectonic faults in the earth's crust, which is in a state of marginal stability. Consequelntly, the results of this work could be used for the discovery of systematic strategies for controlling unwanted seismicity and controlling the complex dynamics that faults and earthquakes display \cite{Turcotte1999}.  It should be mentioned, though, that for the sake of simplicity poroelastic effects are neglected in this work, despite their important role in fault reactivation and induced seismicity \cite{Segall2015}. Their control will be explored in more practical applications of the proposed framework in the future.

\section{Controlling the rich dynamics of Generalized Burridge-Knopoff systems}
\label{sec:numerics}
Generalized Burridge-Knopoff models have rich dynamics and show chaotic behavior. The are also frequently used in statistical physics and geophysics as paradigms of criticality and self-organized criticality. In this Section we first illustrate the rich dynamics of the system by showing its SOC behavior. Then we show how its dynamics can be altered by using the general mathematical developments presented in \ref{sec:control_stabilization} and \ref{sec:control_tracking}, which are based on the mathematical theory of control. The numerical examples show how the system can be stabilized and how it can be driven to stable equilibria of lower (potential) energy in a controlled way avoiding instabilities and cascade phenomena.

\subsection{Self-organized criticality}
\label{sec:SOC}
Many systems in nature show universality in their behavior. A large class of them is believed to exhibit \textit{self-organized criticallity}. These systems are continuously in or close to a state of \textit{marginal stability}, show \textit{chaotic behavior} and obey to similar spatio-temporal correlations and statistical laws. Some examples of SOC are believed to be earthquakes, climate fluctuations, forest and wild fires, snow avalanches, rice- and sand-piles, traffic flows, power electric grids, living organisms (see \textit{Game of Life} \cite{Gardner1969}), population dynamics and evolution, brain neural activity and sparks, stock markets, wars and pandemics \cite<see>[and references therein]{Jensen1998, Turcotte1999, Watkins2016}. 

The term \textit{Self-Organized Criticality} (SOC) was coined in 1988, in the seminal paper of \citeA{Bak1988} \cite<see also>{Bak1989a}, in order to describe the emergence of a critical state in dissipative, dynamical systems, which have no intrinsic time or length scale. The analysis of Bak et al. is based on a cellular automaton \cite{Wolfram1983}, in which a particle is added to a randomly selected cell in a square grid of cells. When a cell in the grid accumulates four particles, the particles are redistributed to their neighboring cells or they are lost (deleted), if they exceed the grid. This conceptually simple model leads to a behavior characterized by long periods of stasis (quiescence) interrupted by intermittent bursts of activity involving the avalanche of few or many particles. These instabilities follow a frequency-area power (fractal) distribution: 
\begin{equation}
N \propto N_f^{-a},
\label{eqn:fractal_distribution}
\end{equation}
where $N$ is the number of avalanches, $A$ the area, i.e. the number of particles involved in the avalanche, and $a\approx 1$. 

Despite the apparent qualitative and statistical similarities in the behavior of many natural systems and numerical idealizations, which are thought to exhibit self-organized critical behavior, the exact definition of the term is somehow unclear. In a recent publication, \citeA{Watkins2016} provide the necessary and sufficient conditions for a system to display SOC, which can be seen as a more a clear definition of the term. These conditions are presented in the discussion (Section \ref{sec:discussionSOC}).
Roughly, SOC implies systems that have the inherent tendency to evolve slowly toward an unstable state, i.e. a critical point. In this sense they are always in or close to a state of marginal stability, without need of external adjustment of their intrinsic parameters.

\begin{figure}
\centering
\includegraphics[width=.7\linewidth]{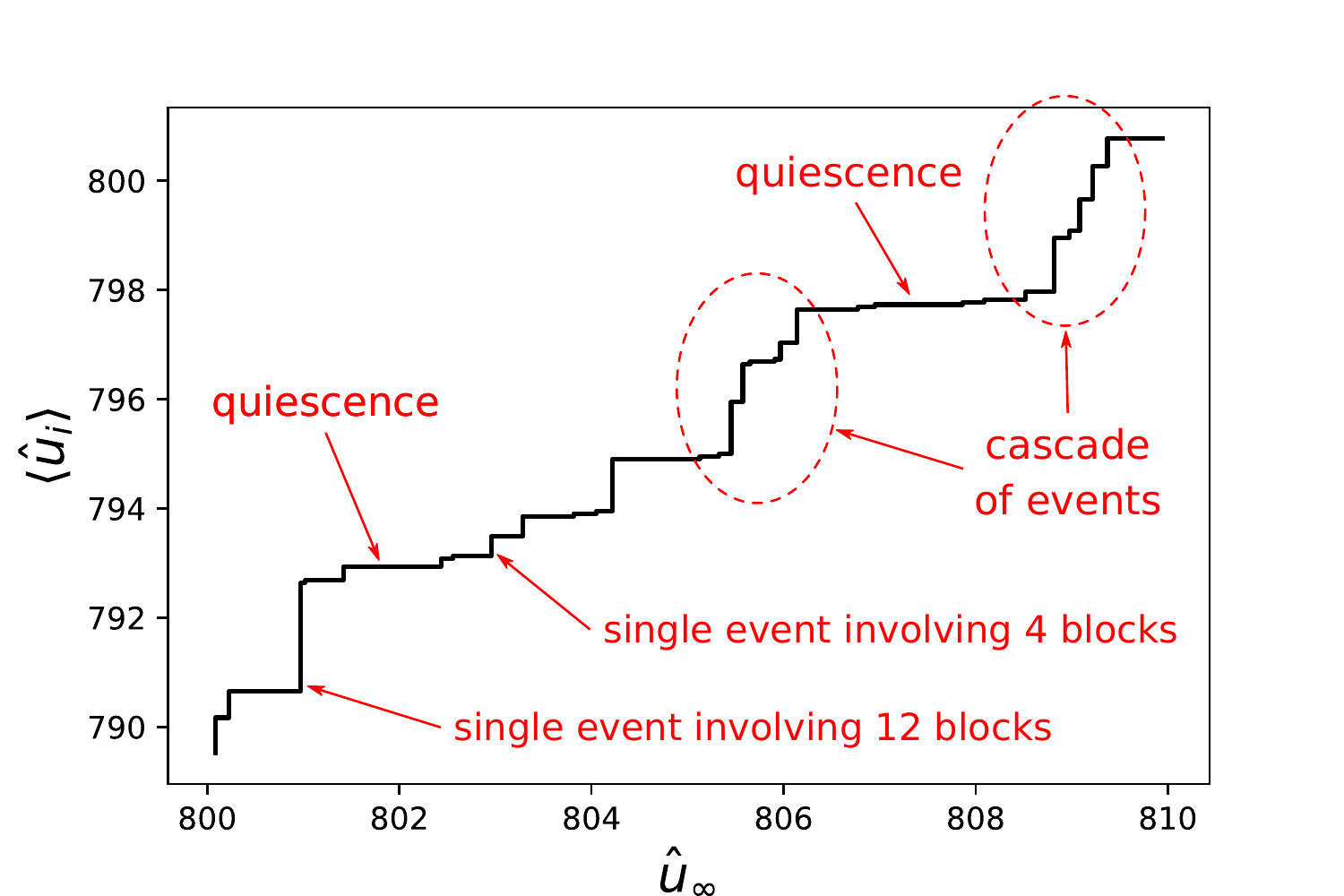} \\*
(a) \\*
\includegraphics[width=.5\linewidth]{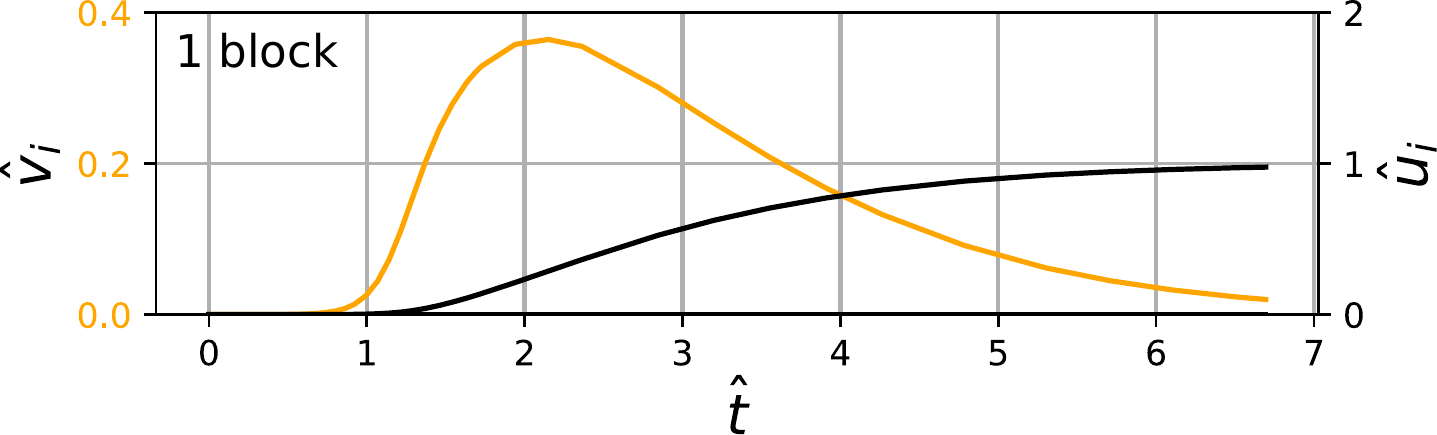} \\*
(b) \\*
\includegraphics[width=.5\linewidth]{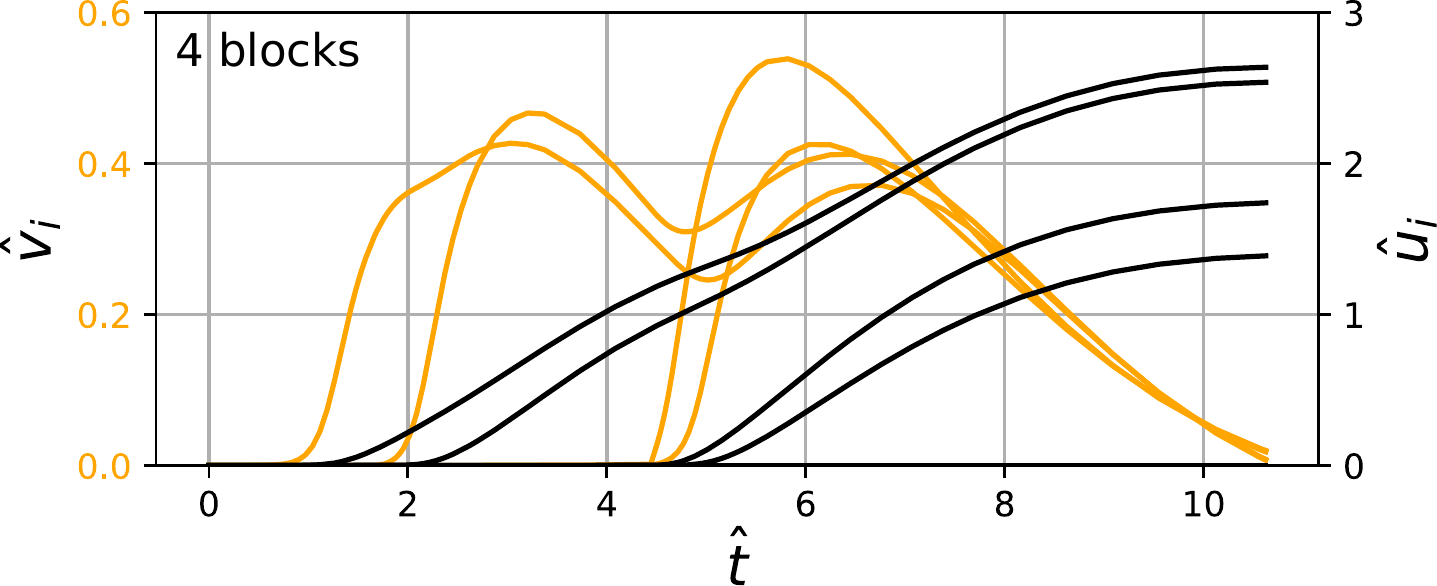} \\*
(c) \\*
\includegraphics[width=.5\linewidth]{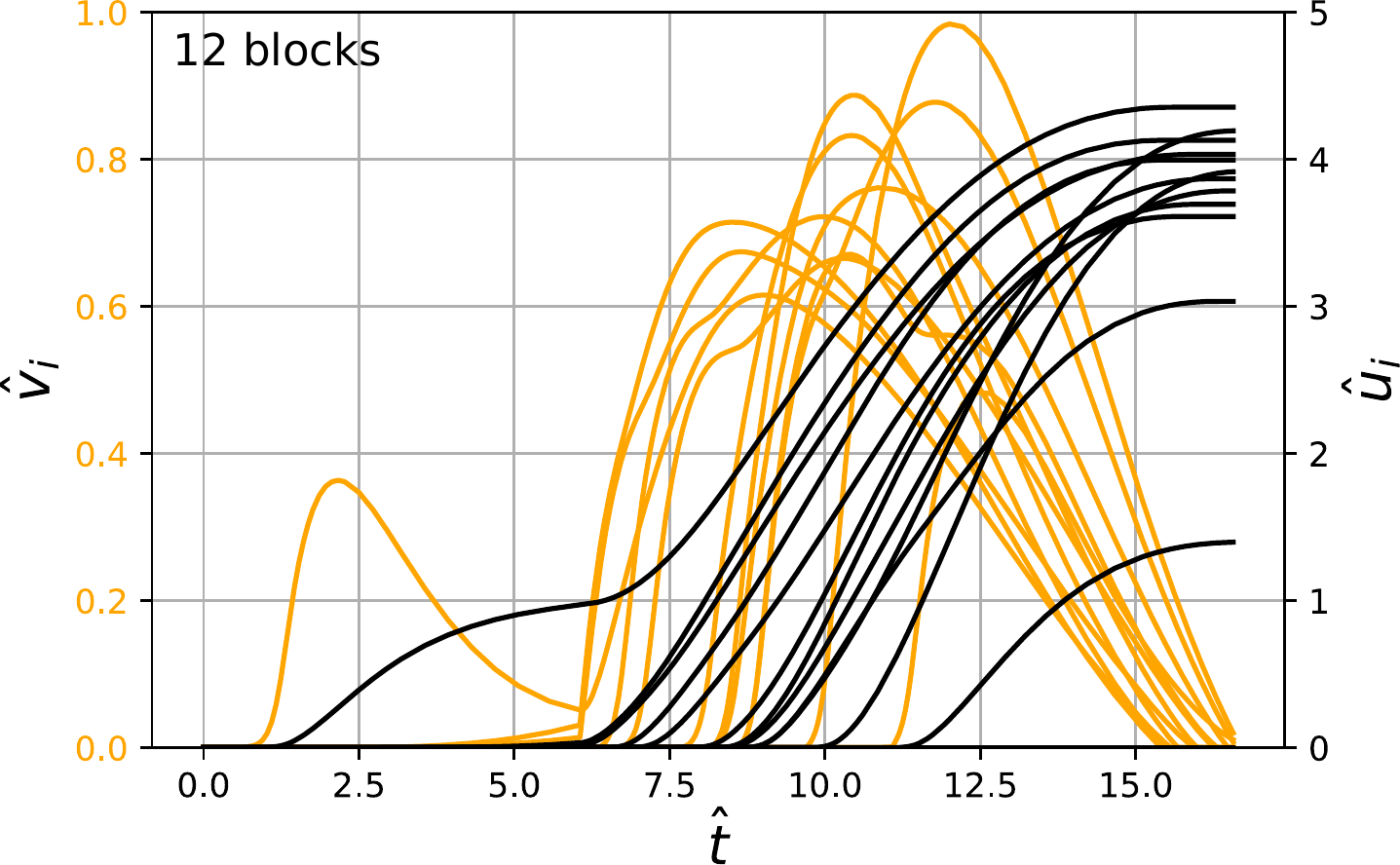} \\*
(d)
\caption{(a) Evolution of average dimensionless accumulated slip, $\langle \hat u_i \rangle=\frac{1}{n}\sum_{i=1}^{n} \hat u_i$, in function of the dimensionless driver plate displacement. The jumps correspond to fast, dynamic events of slip involving one or several blocks (avalanches). Cascade or single events are preceded by large periods of quiescence (plateaus), where energy is accumulated into the system due to the slow movement of the driver plate. (b-d) examples of the sudden, unstable sliding of a single block, of four blocks and of twelve blocks (out of twenty four). Slip, $\hat u_i$, and slip-rates, $\hat v_i$, are reported.}
\label{fig:events}
\end{figure}

Spring-slider models can exhibit self-organized critical behavior \cite{Narkounskaia1992,Turcotte1999}. Due to the continuous motion of the driver plate, they are continuously in or close to a state of marginal stability. In order to illustrate this behavior, we simulate the behavior of a chain of twenty-four (24) blocks. We let our simulations ran until $10000$ events of abrupt sliding are recorded, corresponding to $\hat u_\infty\approx3300$ (details of the simulation procedure and scaling are given in \ref{sec:simulations}). In Fig.~\ref{fig:events}a we present the evolution of the accumulated slip of the blocks (averaged over their total number) with respect to the displacement of the driver plate $\hat u_\infty$. Each step in this plot corresponds to a dynamic event of abrupt sliding of either a single, a cluster or of all the blocks of the system. The magnitude of the observed jumps in displacement depends on the number of the involved blocks in each event. Before an event, a period of quiescence is observed. During this period, the energy is progressively stored in the springs and no slip takes place (see plateaus in Fig.~\ref{fig:events}a). Then the system becomes again unstable and sudden slip occurs, as previously explained. In this sense the system is always in a state of marginal stability.

Figures \ref{fig:events}b-d depict slip events involving a single block and clusters of blocks of various sizes. Often, large events start with the sliding of only one or a couple of blocks, which push their neighbors to sliding in a similar way to a chain reaction (see Figures \ref{fig:events}c-d). The reported slip velocities are high, compared to the slow time scale of the movement of the driver plate. It is worth mentioning that similar behavior is observed in systems with more blocks, in different configurations (cf. Fig.~\ref{fig:gen_slider}) and for real earthquakes. In the case of earthquakes, large events correspond to the main seismic event, while smaller ones to foreshocks and aftershocks \cite{Scholz2002}.

\begin{figure}
\centering
\includegraphics[width=.7\linewidth]{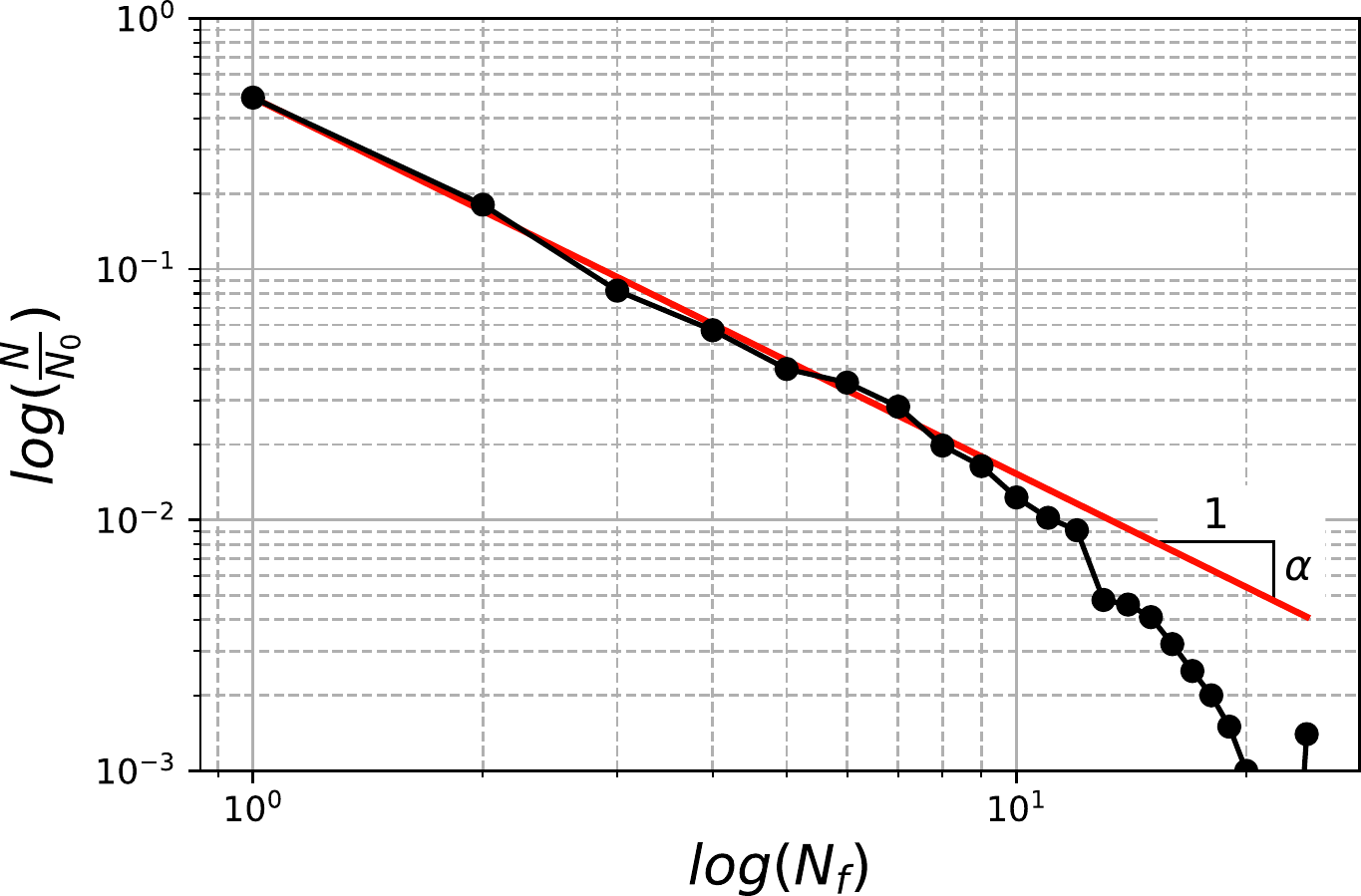}
\caption{Power-law distribution of the frequency ($\frac{N}{N_0}$) - number of blocks involved in slip events ($N_f$) / avalanches. The power law exponent is equal to $a \approx 1.5$.}
\label{fig:powerlaw}
\end{figure}

As shown in Fig.~\ref{fig:powerlaw}, the frequency - number of blocks involved in a slip event ($N_f$) is found to satisfy the power law distribution with $a\approx 1.5$. Similar exponent values were found for different frictional laws and for larger systems of blocks with higher interconnectivity \cite<see>[among others]{Carlson1989,Turcotte1999, Huang1992,Brown1991,Narkounskaia1992,Rundle1991}.  The above power-law distribution can be also transformed to a frequency-event magnitude distribution, which is more common in seismology \citeA<see>[]{Huang1992}. Notice, that the observed divergence from the power-law for events of higher sizes is due to the finite size of the system (here 24 blocks) and it is commonly observed in this kind of simulations in the relevant literature \cite<e.g.>[]{Huang1992}. In this case, the necessary conditions for SOC behavior, which are detailed in the Discussion (Section \ref{sec:discussionSOC}), do not hold and the system is not representative of SOC anymore. Nevertheless, our control strategy is independent of SOC manifestation and is always working independently of the size of the events and the statistics of the uncontrolled system.

Based on the above statistics and the discussion in Section \ref{sec:discussion}, the presented numerical example exhibits chaotic, SOC behavior and can be used for illustrating the efficiency of our control approach. The challenge is therefore to extend this period of quiescence as long as possible and avoid the abrupt energy releases due to sudden sliding. Exploiting the analogy of our system with earthquake faults \cite<see for instance>[\ref{sec:equations_of_motion} and Figure \ref{fig:gen_slider} and Figure \ref{fig:gen_k_weights}]{Rice1993,Bak1989,Turcotte1999}, this would mean that the earthquake instability could be prevented (in theory).

\subsection{Stabilization}
\label{sec:stabilization}
The dynamics of the generalized multiblock system, presented herein, can be controlled using the mathematical theory of control \cite{Vardulakis1991,Vardulakis2012,Ka}. The target is to update the input, which here is the pressure of fluids injected (added) or pumped (removed) at the frictional interfaces of the blocks, in order to stabilize it, i.e. to avoid abrupt slip and sudden energy release. The term stability is used here in the Lyapunov sense (i.e., the system remains close to its equilibrium state under small perturbations from it; for a rigorous mathematical definition of Lyapunov stability we refer to \citeA{Lyapunov1892} and \citeA{Stefanou2016b}).

\begin{figure}[tbhp]
\centering
\includegraphics[width=.7\linewidth]{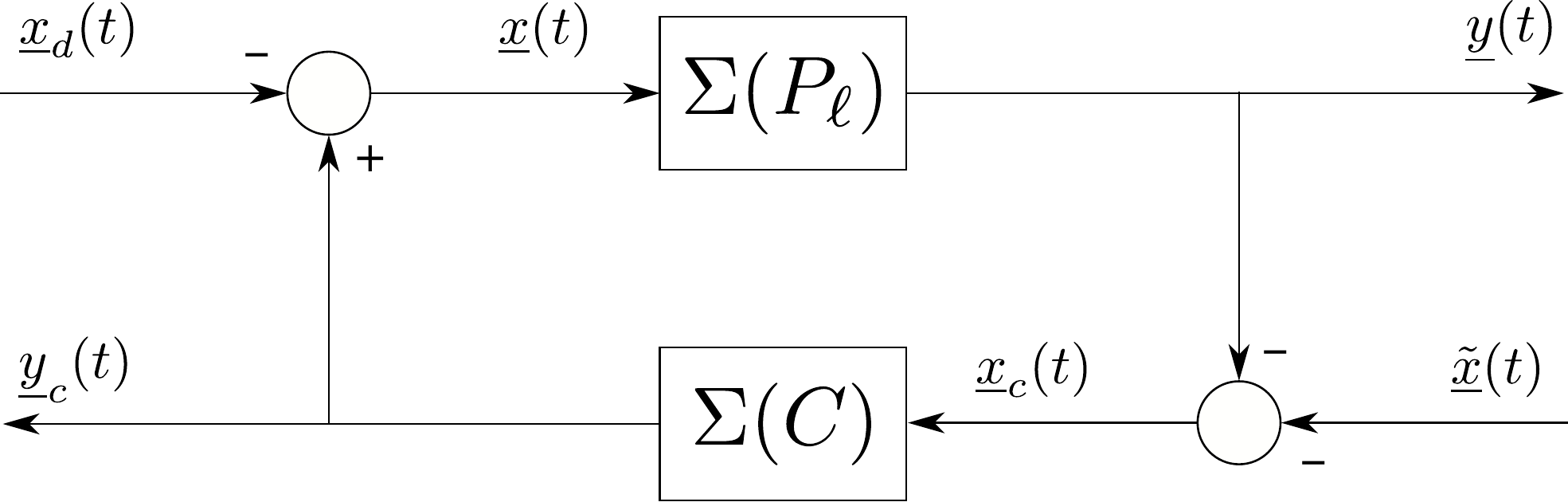}
\caption{Negative feedback control system $\Sigma(P_{\ell},C)$. $\Sigma(P_{\ell})$ is the GBK system (plant) to be controlled with the controller $\Sigma(C)$.}
\label{fig:controller}
\end{figure}

We assume a general \textit{negative feedback} control system as depicted in Fig.~\ref{fig:controller}. $\Sigma(P_{\ell})$ is the multivariable system (plant) to be controlled, i.e. the generalized multiblock system in our case, and $\Sigma(C)$ is the stabilizing controller we need to design. $y(t)$ is the output of the closed-loop, controlled system $\Sigma(P_{\ell},C)$, which here coincides with $\ubar{x}(t)$, $\ubar{y}_c(t)=\ubar{P}(t)$ the output of the controller and $\ubar{x}_c(t)=\ubar{x}(t)+\ubar{\tilde{x}}(t)$ the input of the controller $\Sigma(C)$, with $\ubar{\tilde{x}}(t)$ being a possible perturbation. $\ubar{x}_d(t)$ is a desired state of the system, such that $\displaystyle \lim_{t \rightarrow \infty} \ubar{y}=\ubar{x}_d$. First, we seek the controller $\Sigma(C)$ that can immobilize (or stabilize at the origin in terms of Lyapunov stability) the generalized spring-slider ($\ubar{x}_d(t)=0$). Then we will consider specific forms for $\ubar{x}_d(t)$ (e.g. constant velocity) in order to drive the system smoothly to a desired stable equilibrium point and dissipate the energy in a controlled manner. In the frame of the mathematical theory of control, this process is called \textit{tracking}.

The problem is challenging due to friction and the consequent nonlinearities it introduces. Moreover, the exact values of the frictional parameters are usually unknown. Our stabilizing controller takes into account this uncertainty and is effective even in the absence of complete knowledge of the system's parameters (robustness, \cite{Ka}). As a result, it manages to unravel in real time the unknown, due to uncertainties, dynamics of the system and stabilize it by increasing or decreasing the fluid pressure in the required rate for assuring stability. 

This is achieved using the mathematical developments detailed in \ref{sec:control_stabilization} and \ref{sec:control_tracking}. Our controller guarantees robustness, provided that we can have an estimate a) of the minimum friction coefficient system and b) of the maximum slip and slip-rate softening that can take place. Under this condition, the controller can stabilize and freeze the system at a desired state and consequently prevent sudden sliding. Pressure is adjusted in real time in order to prohibit dynamic events as follows:
\begin{equation}
\ubar{P}=-\frac{1}{2}\uubar{\overline{s}}\uubar{B}^T\uubar{\Theta} \ubar{x}_c,
\label{eqn:control_law_text}
\end{equation}
where $\uubar{B}$ and $\uubar{\Theta}$ are real matrices of constant coefficients and $\overline{\uubar{s}}$ a diagonal matrix containing the sign of the blocks' velocities. These matrices are defined and determined in \ref{sec:control_stabilization}. The presence of the diagonal matrix $\overline{\uubar{s}}$ in the above equation is justified by the fact that the friction is always opposed to the velocity and as such it was included to the GBK model for completeness. Consequently, $\overline{\uubar{s}}$ appears in the control law, even though in practice it is expected to be equal to the diagonal matrix in most cases due to the constant motion of the driver's plate and the subsequent loading of the blocks. Moreover, only tracking under monotonous displacements is of practical interest herein, which implies again that the matrix $\overline{\uubar{s}}$ will coincide with the identity matrix.

The above equation specifies the control law $\Sigma(C)$ and assures global asymptotic stability of the closed-loop system $\Sigma(P_{\ell},C)$, through negative feedback.  Notice, that the above control law is a continuous time regulator and requires continuous monitoring of the state of the system, i.e. of the slip and of the slip-rate. However, in practical situations direct access to the exact state may not be possible. This difficulty could be bypassed by the derivation of observers \cite<e.g.>[]{Ka,Vardulakis1991}, given the observability of our system. Regarding the sampling rate, control can be possible if the system is transformed and studied as discrete-time one. Depending on the technological limitations, the sampling rate could be high or not and could be a parameter to optimize. Quantifying the predictability horizon \cite<cf.>[]{Gualandi2020} of the dynamics of the system could be a useful indicator for optimal control in discrete time.

\begin{figure}
\centering
\includegraphics[width=.7\linewidth]{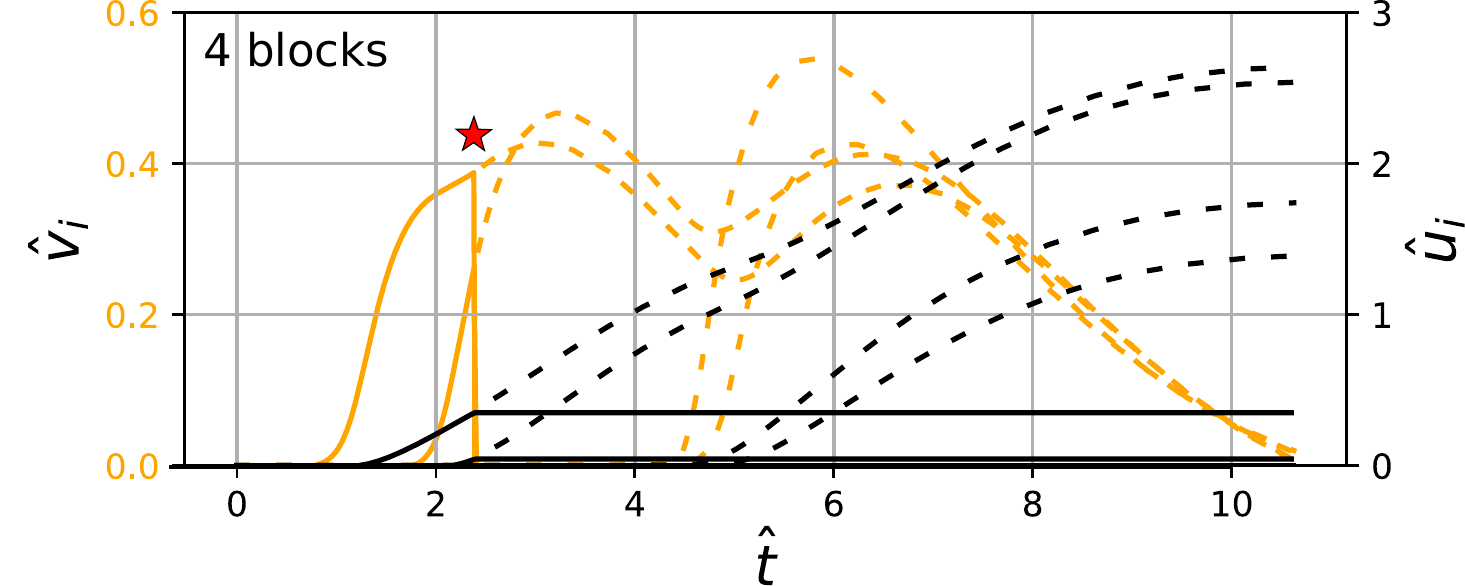} \\*
(a) \\*
\includegraphics[width=.7\linewidth]{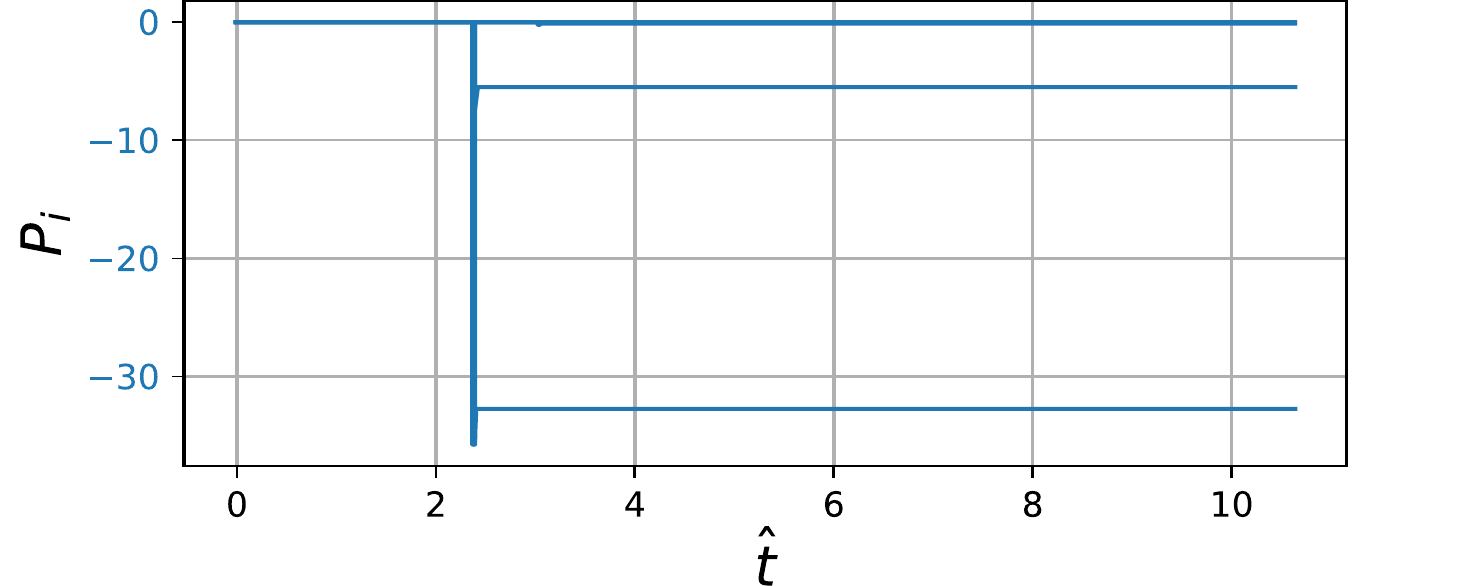} \\*
(b) \\*
\caption{Stabilization of the unstable movement of blocks as presented in Fig.~\ref{fig:events}b. The controller was activated at $\hat t=2.4$ (red star symbol) and successfully stabilized the system by arresting slip (a). Solid lines show the evolution of slip and slip-rate of the sliding blocks before the activation of the controller, while dashed lines show their movement without the controller. The input pressure at the frictional interfaces of the blocks is automatically regulated by the controller (b). Zero $P_i$ corresponds to no fluid pressure change, positive to pumping of more fluid and negative to fluid withdrawal.}
\label{fig:control_instability1}
\end{figure}

Fig.~\ref{fig:control_instability1} shows an example of stabilization of the controlled system. In this example the controller was activated after the initiation of unstable sliding of the two over the four blocks presented in Fig.~\ref{fig:events}c. The controller automatically reduces the fluid pressure at the interfaces of the sliding blocks (Fig.~\ref{fig:control_instability1}b) and immobilizes them as illustrated in Fig.~\ref{fig:control_instability1}a. It is worth mentioning that if the controller is activated before the sliding event, then the instability is completely avoided by tiny decreases in the pressures at the four frictional interfaces of the blocks having the tendency to slide (unstable blocks) and their neighbors. The pressure changes in this case are that small that is not worth of plotting. It is worth emphasizing that this stabilization is achieved without knowing exactly the rheology and frictional properties.

As far as the pressure can be maintained, the system will be stable and no instability (sudden sliding and energy relaxation) will take place. However, if the controller is deactivated, the system will slide abruptly and its unstable character will be restored. Therefore, there is a need to drive the system towards a new, stable equilibrium in a stable, smooth, quasi-static way.

\subsection{Driving the system to lower energy levels without abrupt slip events}

Once a stabilizing, robust controller has been determined, it can be extended in order to make the blocks move to a desired new position with a desired velocity. In the mathematical theory of control, this is called \textit{tracking} \cite{Ka}. The mathematical extension of the controller is presented in \ref{sec:control_tracking}, it has the same form with Eq. (\ref{eqn:control_law_text}) and is robust, meaning that it can drive the system in a controlled manner and in the absence of complete knowledge of its parameters.

In order to illustrate how the controller moves the system to a new position, we focus on the avalanche involving twelve unstable blocks, as presented in Fig.~\ref{fig:events}d. During this event, the total potential energy change was $\Delta \mathcal{E}_{\hat U}\approx -100$. This drop in potential energy happens extremely fast. The maximum velocities reported during the movement of the unstable blocks was of the order of $\hat v\approx1$. Two \textit{control strategies} will be investigated for guarantying at least the same drop in potential energy, but smoothly, without any unstable, uncontrolled movement of any block.

\begin{figure}
\centering
\includegraphics[width=.6\linewidth]{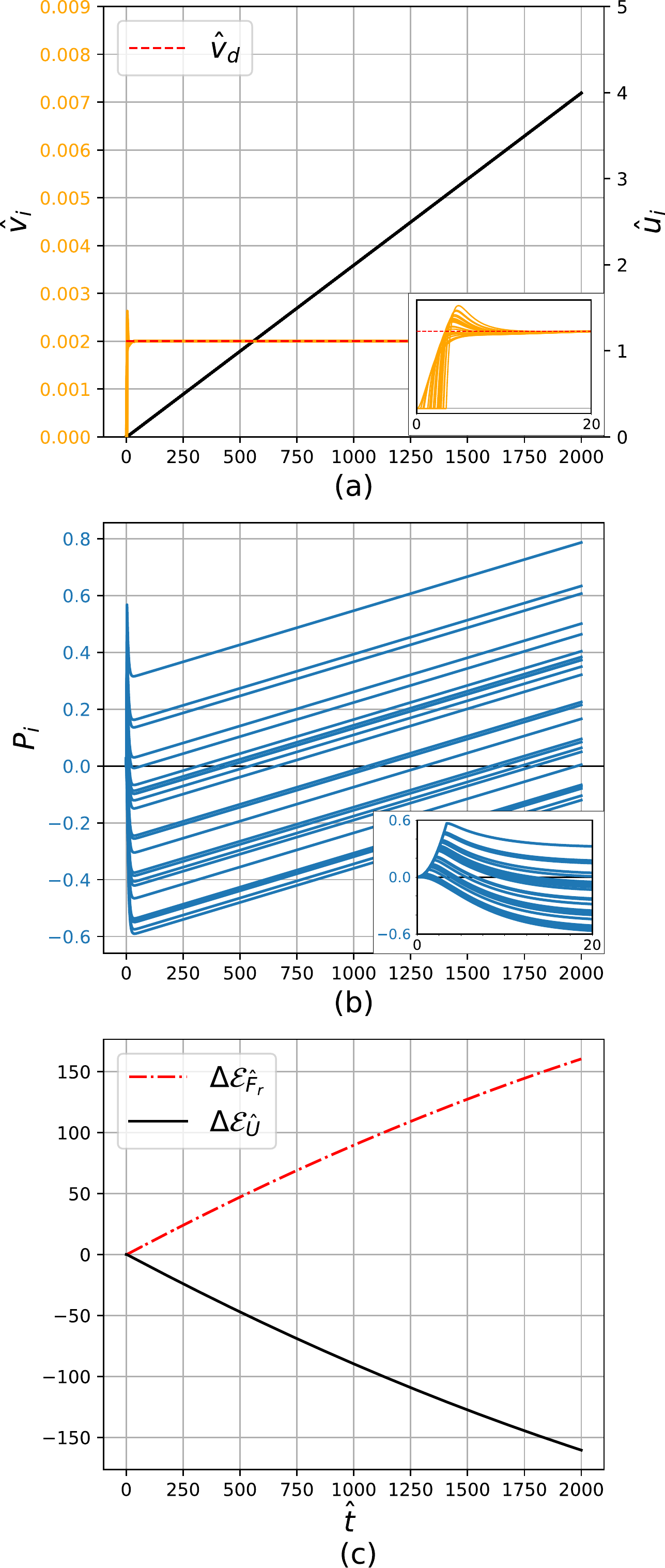}
\caption{First control strategy: time evolution of (a) the displacements, $\hat u_i$, and velocities, $\hat v_i$, of all the blocks, (b) the input pressures, $\hat P_i$, determined by the controller and (c) the total potential energy drop, $\Delta \mathcal{E}_{\hat U}$, and the energy dissipation due to friction, $\Delta \mathcal{E}_{\hat F_r}$. We observe that the controller succeeds in regulating the velocity of all the blocks to the desired value, $\hat v_d$, and dissipate the total potential energy in a controlled and smooth manner, avoiding any instabilities.}
\label{fig:control_tracking1}
\end{figure}

In the first control strategy, the controller will adjust automatically the input pressure in order to assure translation of all the blocks  under the same constant velocity, which is chosen to be equal to $\hat v_i=2\times 10^{-3}$. This target velocity is three orders of magnitude lower than the maximum velocity developed during the unstable movement, but several orders of magnitude higher than the far-field driving velocity, $\hat v_{\infty}$. We allow the system to evolve for a total time equal $\hat t_d=2000$. In Fig.~\ref{fig:control_tracking1}, we present the evolution with time of the displacements and velocities of all the blocks, the input pressure determined by the controller, the potential energy drop and the energy dissipation due to friction. We observe that the controller succeeds in regulating the velocity of all the blocks to the desired value. A small overshoot in velocities at the beginning of the controlled sliding is related to the parameters of the controller chosen (see Eq. (\ref{eqn:control_law4}) in Appendix). This overshoot can be increased or decreased depending on the desired rate of input pressure change. 

\begin{figure}
\centering
\includegraphics[width=.6\linewidth]{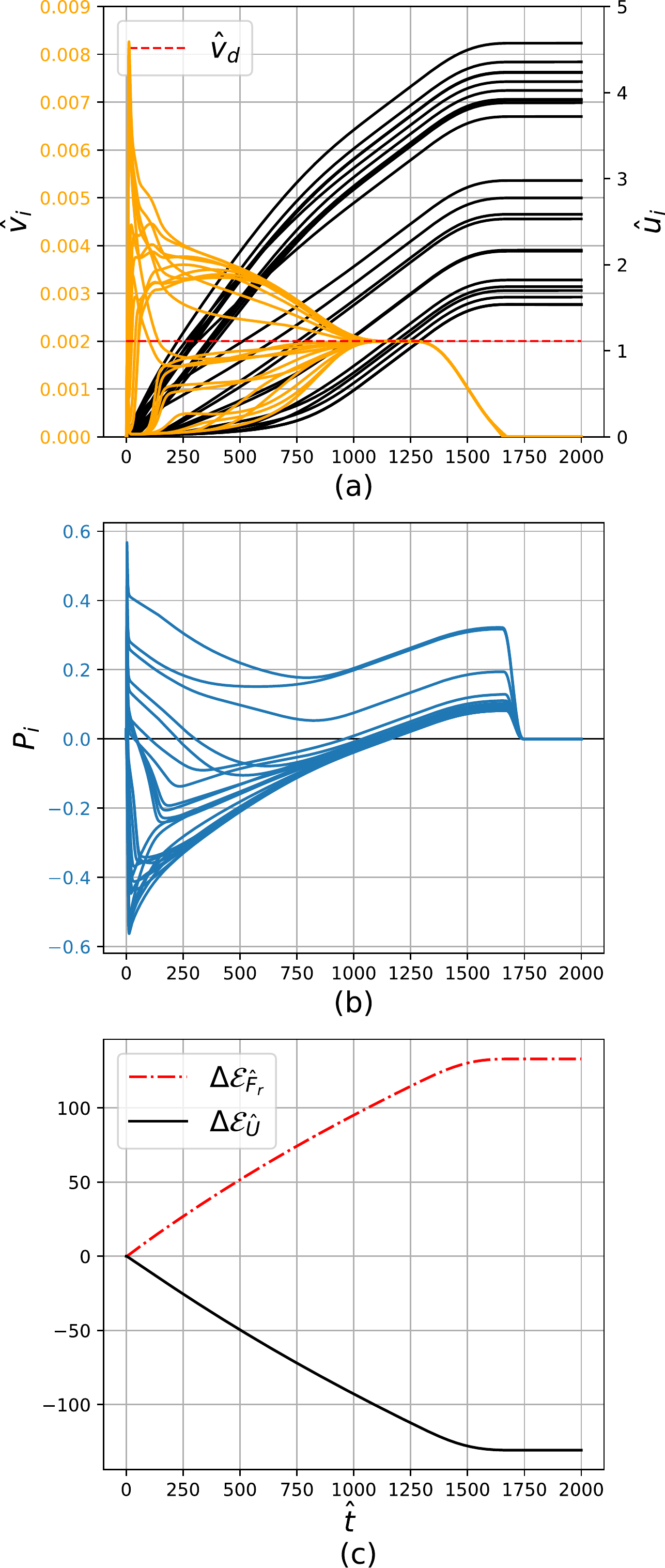}
\caption{Second control strategy: time evolution of (a) the displacements, $\hat u_i$, and velocities, $\hat v_i$, of all the blocks, (b) the input pressures, $\hat P_i$, determined by the controller and (c) the total potential energy drop, $\Delta \mathcal{E}_{\hat U}$, and the energy dissipation due to friction, $\Delta \mathcal{E}_{\hat F_r}$. We observe that the controller succeeds in regulating the average velocity of the blocks to the desired value, $\hat v_d$. The blocks self-organize to a desired stable equilibrium state. Then, after $\hat t=1250$, the controller is progressively deactivated. Any instabilities are prevented and the total potential energy is dissipated efficiently, as desired.}
\label{fig:control_tracking2}
\end{figure}

At the beginning the applied pressure change is zero, but the controller adjusts it automatically, in order to allow for the blocks to attain the desired velocity, as shown in Fig.~\ref{fig:control_instability1}b. Two groups of blocks are distinguished, i.e. those that are on the verge of unstable movement and those that are in a stable equilibrium. In the case of the former, we observe that the input pressure is decreased (negative pressure change; fluid withdrawal). As a result the regulator increases friction, decelerates the movement of the blocks and stabilizes the system. On the contrary, in the case of the latter, the controller increases the input pressure in order to accelerate their motion and achieve the desired target velocity. The pressure range (ordinate in Fig. \ref{fig:control_tracking1}b), determined automatically by the controller, depends on how far from equilibrium each block is, while the evolution of pressure with time (abscissa  in Fig. \ref{fig:control_tracking1}b) on the target velocity, $\hat v_i$.

It is worth emphasizing that the regulator is not based on any ``if-then'' statements - this would be impossible in general situations. On the contrary, based on the mathematical developments presented in the appendix, it automatically regulates the fluid pressure of the blocks and stabilizes the system by unraveling its dynamics in real time. This is accomplished by monitoring the motion of the system and adjusting the input pressures changes in real-time. Moreover, it has to be mentioned that the regulator is agnostic to the exact frictional parameters and frictional rheology.

Regarding the blocks that started from a stable equilibrium, at a certain point they also enter to a critical state due to slip accumulation. At that point, the controller decreases their input pressure, in order to alter the dynamics of the system, guarantee stability and achieve motion under the desired velocity. This is depicted in Fig.~\ref{fig:control_tracking1}b by the negative pressure change at all the blocks (pressure reduction).

The stored energy in the system is dissipated almost linearly with time as shown in Fig.~\ref{fig:control_tracking1}c. The same holds for the potential energy decrease in the system. Notice that thanks to the desired very low target velocities, the kinetic energy of the system and the viscous dissipation are negligible compared to the drop in potential energy. With our approach, we finally manage to dissipate more than the released energy during the unstable, abrupt movement of the system, but in a slow, smooth and controlled manner. Consequently, the system is not anymore in a state of a marginal stability and it cannot present a self-organized critical behavior \cite<or richer, see>[]{Ben-zion2008}. Moreover, as its motion is actively controlled, no chaotic behavior can be observed. In other words, we managed to completely alter the dynamics of the system in a desired way.

However, the results show that the controller has to assure a negative input pressure in order to prevent the unstable movement of some blocks (see Fig.~\ref{fig:control_tracking1}b). These lower pressures than the initial ones have to be maintained for assuring stability. Therefore, if we decided to deactivate the controller, the system would become unstable again. Nonetheless, it is possible to bring it in a state of stable equilibrium, in which the controller is no more necessary and can be deactivated. For this purpose, we follow a different scenario and instead of setting the velocity of each one of the blocks to the desired velocity, we set the average velocity equal to $\langle \hat v_i \rangle=\frac{1}{n}\sum_{i=1}^{n} \hat v_i=2\times 10^{-3}$. In this manner we incite a faster, but controlled movement for the unstable blocks for which the input pressure is negative. As a result some blocks move faster than others, noting higher displacements (Fig.~\ref{fig:control_tracking2}a). After some time we observe that the blocks self-tune and slide with identical slip-rates equal to the desired one. Once more, the system's dynamics were altered and no self-organized critical or chaotic behavior occurred. On the contrary, we incited the system to self-organize towards a stable equilibrium (see $\hat t>1200$ in Fig.~\ref{fig:control_tracking2}). This is indicated by the positive input pressures changes at the frictional interfaces of all the blocks, which are required to sustain the movement under the desired average velocity. 

In order to illustrate the stability of the system in this new state, we decide to progressively deactivate the controller between $\hat t=1250$ and $1750$, as depicted in Fig.~\ref{fig:control_tracking2}. Notice, that the system remains now in a state of stable equilibrium of a lower (potential) energy level. Of course, if the controller remains inactive, the continuous slow movement of the driver plate will render again the system unstable after a (large) time interval. Therefore, it might be interesting after this point to set the controller's target velocity equal to the driver plate's velocity. In this case the regulator will automatically adjust the fluid pressure and the blocks will follow the movement of the driver plate sliding continuously in an aseismic way and by small pressure changes.

The above numerical examples show that the system is finally stabilized by reducing the pressure at the blocks' interfaces and not by increasing it. This might be at first counter-intuitive, if stabilization is thought as a simple process of adjusting pressure in order to satisfy the frictional stability condition of the system. For the single spring-slider model, frictional stability is guaranteed when the stiffness of the loading system is higher than a critical softening stiffness, which depends on the exact frictional parameters and rheology \cite<see >{Ruina1983,Scholz2002,Stefanou2019}. For the GBK model, the stability conditions are qualitatively similar. Nevertheless, attempting to stabilize the system by satisfying the frictional stability conditions (provided that the fault is reactivated) would require very high positive fluid pressure changes. For typical values of stiffness and frictional weakening parameters of faults the pore fluid increase should be close to the insitu effective stress. Moreover, stabilizing the system in that way wouldn't mean that it would be immobilized. On the contrary it would slide aseismically,  but uncontrollably. In the best case it would follow the far-field tectonic velocity. Moreover, very high velocities and instabilities could be developed during the pore pressure increase after fault reactivation.

However, with our approach we achieve stabilization with minimal pressure changes. Additionally, we guarantee controlled sliding, with a desired target velocity profile, decided by the operator. This can be quite important for industrial projects involving fluid injections in depth. Some preliminary, practical numerical examples on the base of the simple spring-slider model were given in \citeA{Stefanou2019}, where pressure reduction is found again to be needed for stabilization. In these examples tracking was achieved by increasing the effective stress by $\sim 15\%$. Similar practical examples for more complicated systems can be given with the proposed approach, but this extends the scope of the present work, which has a more theoretical character. Furthermore, no bounds to the pressure change and/or the pressure-rate change were set in this analysis. Depending on the exact technological and engineering constraints, limits on the pore pressure increase/decrease could be imposed using techniques from the mathematical theory of control. Optimal control could be also designed based on engineering criteria for specific applications.

As far as it concerns fluid injections in real scale applications, there are several examples in the literature correlating induced seismicity with fluid pressure increase (see above cited works, among many others). Of course, correlation is not causality, especially in complex systems as the one at hand and several interpretations might exist. Our numerical results show that fluid pressure increase leads to slip acceleration and fluid pressure decrease to slip arrest. This is justified from the physics point of view (decrease vs increase of effective stress and therefore of friction). Moreover, it seems to be corroborated with fluid injections and production at the low temperature geothermal field Laugaland \`{i} Holtum in the south Iceland Seismic zone. According to \citeA{Flovenz2015} the decrease of the fluid pressure due to geothermal production (fluid withdrawal) and seasonal variation in the pressure seem to have modulated the natural seismicity by delaying an impending $M_w=6.4$ event on June 17th 2000 by several years and affecting its exact timing. Pursuing further the proposed methodology and if the current level of technology allows, we might be able one day to mitigate such events.

\section{Discussion and concluding remarks}
\label{sec:discussion}
\subsection{Self-organized criticality control}
\label{sec:discussionSOC}
SOC could be seen as a spectacular manifestation of order in nature that results in sparks of energy relaxation (dissipation). Nonetheless, this does not mean that SOC behavior cannot be prevented. The necessary and sufficient conditions for SOC that were recently proposed by \citeA{Watkins2016} and are presented below, leave open this possibility.

\textit{Self-Organized Criticality Control} (SOCC) can be of particular importance in many situations where avalanches due to SOC behavior are unwelcome. SOCC is a relatively new field. Maybe the most popular example of SOCC is the prevention of large snow avalanches, by triggering smaller ones \cite{McClung1993,Birkeland2002}. \citeA{Cajueiro2010a,Cajueiro2010,Cajueiro2010b} applied and extended this idea for controlling self-organized criticallity in the Abelian sand pile model \cite{Dhar1989} and generalizations of it. \citeA{Brummitt2012} studied the suppression of cascade failures in interconnected powergrids, based on the sand pile model of \citeA{Bak1988}. Again using as model the classical sandpile automaton of Bak et al., \citeA{Noel2013} proposed a control strategy that determines the grid cell in which a particle should land in order to adjust the probability of triggered cascades and mitigate large avalanches. Another example of SOC control is given by \citeA{Hoffmann2014}, who altered the SOC power law statistics of electrical circuits obeying Kirchoff's law, by adequately modifying the interconnectivity of the circuit network. In this way they proposed mitigation strategies of large cascade events.

According to \citeA{Watkins2016}, a system has to satisfy the following three \textit{necessary} conditions in order to qualify as SOC:
\begin{itemize}[noitemsep,topsep=0pt]
  \item [NC1.] Non-trivial scaling.
  \item [NC2.] Spatio-temporal power law correlations.
  \item [NC3.] Apparent self tuning to the critical point.
\end{itemize}
These necessary conditions are considered in the logical sense. In other words, a system cannot exhibit SOC if any of the above three conditions is not fulfilled.

An extensive discussion of the meaning of critical point and criticality in the frame of SOC and its connection with existing notions in physics and statistical mechanics is made in \citeA{Watkins2016}. Here, critical points are points in the phase portrait of the system that are (Lyapunov) unstable. Indeed, due to the slow driver plate's movement the system is evolving continuously toward a critical, unstable equilibrium followed by abrupt cascade events (non-equilibium states in the mathematical sense, see also Figure \ref{fig:events}a). These events may be small, involving few blocks or large, involving several blocks.

\citeA{Watkins2016} give also the following three \textit{sufficient} conditions for characterizing a system as SOC:
\begin{itemize}[noitemsep,topsep=0pt]
  \item [SC1.] Non-linear interaction, normally in the form of thresholds.
  \item [SC2.] Avalanching.
  \item [SC3.] Separation of time scales.
\end{itemize}
That means that if a system fulfills these conditions, then it can exhibit SOC.

The open-loop, uncontrolled GBK model, $\Sigma(P_{\ell})$, satisfies all the sufficient conditions and therefore exhibits SOC. In particular,  friction introduces the necessary non-linearities and, due to slip or slip-rate weakening, it takes a maximum value before slip initiation (threshold). Avalanches are also observed involving clusters of blocks that dissipate abruptly the energy of the system (intermittent energy relaxation). The driver plate's slow movement introduces a slow time scale in $\Sigma(P_{\ell})$ (see time-scale asymptotic analysis in \citeA{Stefanou2019}), while the events follow the fast  characteristic times related to the frictional instability during avalanches. 

In contrast, the controlled, closed-loop system, $\Sigma(P_{\ell},C)$, remains strongly non-linear, satisfying condition SC1, but not conditions SC2 and SC3. Moreover, it does not satisfy the necessary condition NC3, because our controller, $\Sigma(C)$, is designed to prohibit self-tuning to a critical point. Instead, the closed-loop system is self-tuned toward desirable stable equilibria. As a result, based on the necessary conditions of \citeA{Watkins2016}, the controlled system cannot exhibit SOC. Self-organized criticality is controlled.

It is worth pointing out that our control approach differs from the aforementioned SOCC approaches in many aspects. First of all, it is based on a totally different mathematical framework. This framework allows us to derive rigorous mathematical proofs about the stabilization and controllability of the non-linear system. Moreover, it allows to alter its dynamics by considering also the uncertainties of the physical model (robustness) in a deterministic way. Notice that the numerical examples presented in Section \ref{sec:numerics} are only for illustrating the mathematical findings and not for proving or verifying them. Another different aspect of our approach is related to the underlying model and the chosen input. Even though sandpile systems and networks exhibit SOC and display rich dynamics, they differ from the GBK frictional model considered herein. GBK is described here by a set of Ordinary Differential Equations and it is prone to the application of the mathematical tools of control theory. Moreover, we don't trigger any instabilities to dissipate energy, as it is done in the above cited works, and we do not artificially increase locally the energy of the system or change its interconnectivity by using statistical methods. In this sense our approach is deterministic even though it considers the uncertainties of the underlying physics. As a result we go beyond existing SOCC approaches by slowing down the dynamics of the GBK model. In this way, we do not only dissipate the required energy in a controlled manner, but we also break the separation between slow and fast dynamics of the system, which is key ingredient for SOC as stated above. 

\subsection{Restriction of chaotic behavior}
\label{sec:discussionCHAOS}

Chaotic behavior is also restricted. While the evolution of the uncontrolled GBK system is chaotic and could be predictable only in a statistical sense, due to complexity and chaos, the evolution of the controlled system is not. The presence of the controller guarantees global asymptotic stability (see \ref{sec:control_stabilization}). Hence no limit cycles or chaos are possible \cite{Strogatz1994}. Moreover, the controller is robust, meaning that it succeeds in altering the dynamics of the system even without knowing its exact properties. Indeed, only some rough boundaries of the frictional parameters are needed in order to drive and control the system as desired.

\subsection{Implications for complex (geo)systems}

Following the seminal work of P. Bak and his colleagues in 1988 \cite{Bak1988}, a broad range of natural, technological, biological and social phenomena were identified to show self-organized critical behavior. Some scientists will claim that Self-organized criticality is ubiquitous in nature. However, some others will be skeptic and will criticize the universality of self-organized critical behavior \cite<see discussion in>[]{Watkins2016}. The latter will be based on the limitations of available data and observations. Further research is needed before adopting one or the other side. For a debate on the applicability of the SOC concept to earthquakes we refer to \citeA{Ben-zion2008,Sornette1999,Main1999,Lomnitz-Adler1993}, among others.

Nevertheless, what is beyond any doubt, is that real-world systems do present cascade failures. The failure of a node in a multi-node system (here, of a block of the GBK system) can trigger accelerating feedback and cause the failure of other nodes in the system, in a \textit{domino-like} way. The reason is that systems in nature are inherently complex and allow information to spread in several spatio-temporal scales. This occurs in a way that it can be difficult or even impossible to grasp and model in details. 

Here, we present an example which, with a quite high degree of abstraction, shows that cascade failures could be prevented, even if we are agnostic to the details of the exact spatio-temporal correlations and dynamics of the underlying system. In other words, a complete understanding of the physics behind cascade failures and of the interconnectivity of the system's nodes is not an absolute requirement for preventing avalanches. This is proved mathematically and illustrated through some examples for the strongly non-linear, complex geophysical system at hand, by assuring robust control. 

Furthermore, having the possibility of controlling the dynamics of a complex system in a robust way, can give us useful information about its inherent, but practically inaccessible properties. Here, we drive smoothly the system to desired stable equilibrium states, which, though, are a priori unknown. By back-analyzing the evolution of the stabilizing input, one could draw real-time conclusions about the interconnectivity, the dynamics and the evolving hidden characteristics of the system. Therefore, the proposed strategy could help in improving the current understanding in some systems that complexity makes opaque.

An additional implication of the proposed theory has to do with \textit{predictability}. Predicting the evolution of complex systems exhibiting self-organized critical behavior (or richer) is a challenging, but controversial topic \cite<see>[for an overview]{Watkins2016, Ben-zion2008, Sornette1999}. Nevertheless the ability to predict, can have important consequences in many disciplines (cf. earthquakes, tectonics, volcanoes). The distribution and frequencies of cascade events of the model presented herein or of more complex ones is a useful statistical correlation. However, it cannot provide with certainty when and how large exactly the next cascade failure will be. Nevertheless, if control is possible, as shown here, then prediction is irrelevant. The more we control a system, the less unpredictable it becomes. Of course, one would need sufficient inputs and monitoring for guaranteeing full control \cite<see \textit{controllability} and \textit{observability} notions of the mathematical Theory of Control in>[among others]{Ka}. However, even if only partial control is possible (e.g. due to limited area of intervention and technological constraints), the space of uncontrolled dynamics will be reduced, which can lead to improved predictability (e.g. increase the predictability horizon \cite<e.g.>[]{Gualandi2020}) and constrain the size of the next cascade failure.

\subsection{Implications for anthropogenic and natural earthquakes}

A direct implication of the present work is inevitably related to earthquakes. Burridge-Knopoff models are frequently used as qualitative analogs of the earthquake phenomenon, either at the level of single fault or of complex fault networks. 
Of course, models of the Burridge-Knopoff type have several limitations as far it concerns the representability of the earthquake phenomenon. These limitations are well identified in the early work of \citeA{Rice1993} and relevant literature. However, the mathematical structure of the generalized Burridge-Knopoff models allows the development of a general control approach that could be applied in more realistic situations of earthquake rupture, provided that a consistent fault discretization approach is followed \cite<see>[among others and relevant numerical methods in elastodynamics]{Rice1993,Chinnery1963,Erickson2011,Ben-zion2008}. Therefore, more realistic cases could be tackled using the theoretical developments presented herein (see \ref{sec:equations_of_motion}, Figure \ref{fig:gen_k_weights} and Figure \ref{fig:gen_slider}), but this extends the scope of the present work.

Moreover, it is worth emphasizing that we show a way of active stabilization of the generalized Burridge-Knopoff without knowing its exact properties. This means that detailed knowledge of faults' frictional parameters, which is practically impossible to acquire in practice, might not be a \textit{sine qua non} condition. Notice, that friction is considered as the cornerstone for understanding earthquake behavior and it is a major unknown \cite{Erickson2011}. However, our control approach needs minimal and not precise information about the frictional characteristics of the fault system, which can be easily acquired in practice. Based on this limited information, we show how the system can be driven to a stable state in a totally controlled and aseismic way. Moreover, our approach guarantees aseismic, slow-slip and smooth energy relaxation and does not require the knowledge of the exact current stress state and tectonic setting. The system is controlled independently of being far or close to its critical points.

Without any doubt, claiming that controlling anthropogenic or natural seismicity is possible, based on the analysis presented herein, is a speculation and further research is needed. Several theoretical and techno-economical investigations have to be pursued further in order to show into what extend man-made or natural earthquakes can be prevented (or the opposite). For example, some direct limitations of the proposed theory have to do with the actual technological means for fluid injections in the earth's crust, the sampling-rate and frequency of observations, the in-situ hydrogeological and geomechanical conditions and uncertainty quantification. However, the current work sets the mathematical and physical framework for inspiring further research on controlling induced seismicity and, maybe, at a later phase, on controlling natural seismicity as well.

\section*{Acknowledgements}
I would like to thank J.-P. Avouac for his insightful feedback, constructive comments and our fruitful discussion regarding the application and the limitations of the proposed methodology and model. I would like also to thank A. Gualandi for his constructive criticism and positive feedback.\\
This work was supported by the European Research Council (ERC) under the European Union Horizon 2020 research and innovation program (Grant agreement 757848 CoQuake), \url{http://coquake.eu}.


\appendix

\section{Equations of motion}
\label{sec:equations_of_motion}
The equation of motion of block $i$ is written as follows:
\begin{equation}
\begin{split}
m_i \dot v_i&=
\sum_{j=1}^{n} k^c_{ij} \left(u_j-u_i\right)
+\sum_{j=1}^{n} \eta^c_{ij} \left(v_j-v_i\right) \\
&+k^l_{i}\left(u_{\infty}-u_i\right)+
\eta^l_{i}\left(v_{\infty}-v_i\right) \\
&+\sum_{j=1}^{n} k^c_{ij} \left( u^0_j-u^0_i\right)-F^r_i,
\end{split}
\label{eqn:block_i}
\end{equation}
where $\dot{(.)}$ is the time derivative, $u_i$ and $v_i$ are, respectively, the slip (displacement) and slip-rate (velocity) of the block $i$, $u_\infty$ and $v_\infty$ are, respectively, the displacement and velocity of the driver plate, which represents the far field tectonic velocity in the case of faults, and $u^0_i$ is the initial displacement of the block $i$. $k$ stands for stiffness and $\eta$ for damping coefficients. The superscript `$c$' denotes the springs and dampers between the blocks and the superscript `$l$' the same elements between the blocks and the driver plate. For instance, $k^c_{ij}$ is the stiffness coefficient of the spring connecting block $i$ with block $j$, while $\eta_i^l$ is the damping coefficient of the dashpot connecting the block `$i$' with the driver plate. $F^r_i$ represents the friction of block $i$ with the rough plane and can depend on slip, rate of slip and other internal state variables (see section \textit{Friction and instability}, \ref{sec:friction}). Here $F^r_i=F^r_i\left(u_i,v_i\right)$ (see Fig.~\ref{fig:gen_slider}d,e). \\
Setting $\omega^*=\sqrt{\frac{k^*}{m^*}}$, where $k^*$ and $m^*$ are, respectively, a reference stiffness and mass, the above equations take the dimensionless form:
\begin{equation}
\begin{split}
\hat{m}_i \hat{v}'_i&=\sum_{j=1}^{n} \hat{k}^c_{ij} \left(\hat{u}_j-\hat{u}_i\right)
+\sum_{j=1}^{n} 2 \zeta \hat \eta^c_{ij} \left(\hat v_j-\hat v_i\right) \\
&+\hat k^l_{i}\left(\hat u_{\infty}-\hat u_i\right)+
2 \zeta \hat \eta^l_{i}\left(\hat v_{\infty}-\hat v_i\right) \\
&+\sum_{j=1}^{n} \hat k^c_{ij} \left( \hat u^0_j-\hat u^0_i\right)-\hat F^r_i,
\end{split}
\label{eqn:dim_block_i}
\end{equation}
where $(.)'$ is the derivative with respect to the dimensionless time $\hat t=\omega^* t$, $\zeta=\frac{\eta^*}{2 m^* \omega^*}$ is the damping ratio, $\hat k^{c,l}_x=k^{*-1} k^{c,l}_x$, $\hat \eta^{c,l}_x=\eta^{*-1} \eta^{c,l}_x$, $\eta^*=2 \zeta m^* \omega^*$, $\hat u_i=D^{*-1} u_i$, $\hat v_i=v_i D^{*-1} \omega^{*-1}$, $D^*$ a reference displacement and $\hat F^r_i=(D^* k^*)^{-1} F^r_i$.

Equations \eqref{eqn:dim_block_i} are written in matrix form as follows:
\begin{equation}
\ubar{x}'= \underbrace{\begin{bmatrix}
\uubar{O} & \uubar{I} \\
-\uubar{K} & -2\zeta \uubar{K}
\end{bmatrix}}_{\uubar{G}} \ubar{x} + \underbrace{\begin{bmatrix}
\ubar{O} \\
\ubar{\Psi}
\end{bmatrix}}_{\ubar{H}}.
\label{eqn:dynamics}
\end{equation}
The first $n$ components of the vector $\ubar{x}$ represent the dimensionless displacements and the rest $n$ components the dimensionless velocities of the blocks. $\uubar{O}$ and $\uubar{I}$ are, respectively, the zero and identity matrices of size $n \times n$ and $\ubar{O}$ the zero vector of size $n$. $\uubar{K}=\uubar{M}^{-1} \left(\uubar{K}^l-\uubar{K}^c \right)$, where $\uubar{M}$ is a diagonal matrix containing the dimensionless masses of the blocks, $\{\uubar{M}\}_{ii}=\hat m_i=1$, $\{\uubar{K}^c\}_{ij}=\hat k^c_{ij}$ and $\uubar{K}^l$ a diagonal matrix with components $\{\uubar{K}^l\}_{ii}=\hat k^l_{i}$. We say that the system is in equilibrium when $\ubar{x}'=0$. In the above equation we assumed that the dimensionless damping coefficients coincide with the dimensionless stiffnesses. This is a reasonable assumption in the absence of more detailed data. The vector $\ubar{\Psi}$ represents the dimensionless forces applied to the blocks due to the initial deformation of the springs, $\hat u_i^0$, the displacement, $\hat u_{\infty}$, and velocity, $\hat v_{\infty}$ of the driver plate, and the friction forces $\ubar F^r$: $\ubar\Psi=\uubar{M}^{-1}\left( -\ubar{F}^r-\uubar{K} \ubar{U}^0 +\uubar{K}^l \ubar{U}^{\infty}+2\zeta \uubar{K}^l \ubar{V}^{\infty}  \right)$, where $\{\ubar F^r\}_i=\hat F^r_i$, $\{\ubar U^0\}_i=\hat u^0_i$, $\{\ubar U^\infty\}_i=\hat u^\infty_i$ and $\{\ubar V^\infty\}_i=\hat v^\infty_i$.

\begin{figure*}
\centering
\includegraphics[width=1.\linewidth]{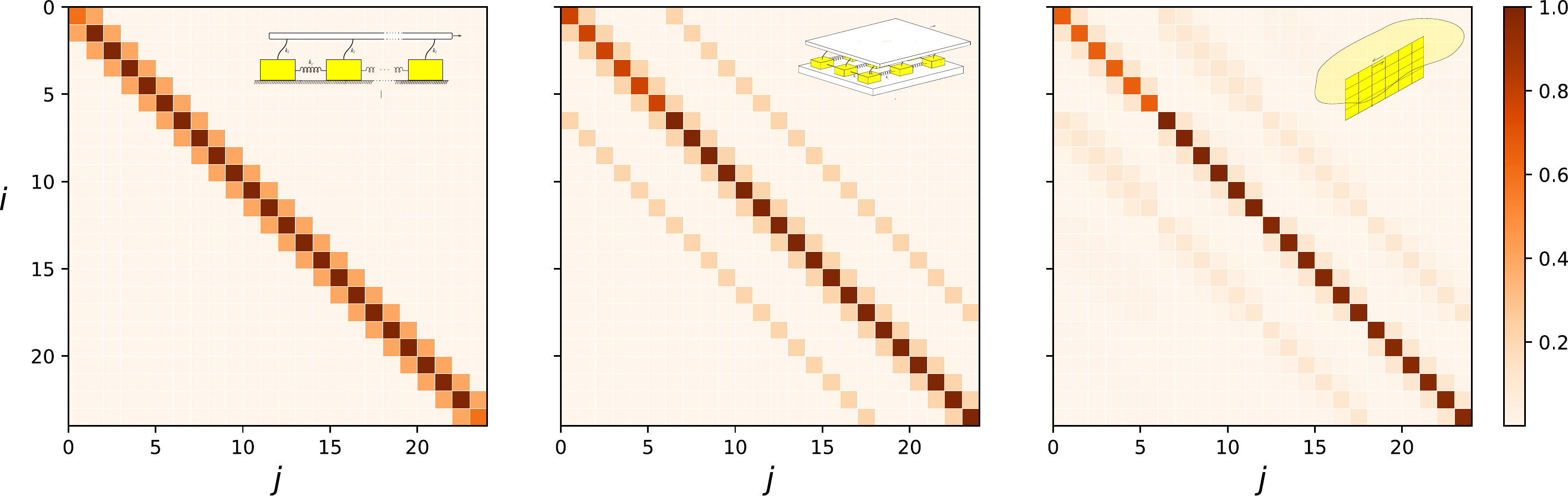}
\caption{Graphical representation of the $(i,j)$ components of the connectivity matrix, normalized by its maximum diagonal component, for the Burridge-Knopoff model with $24$ blocks, its two-dimensional generalization with a $4 \times 6 = 24$ blocks and a strike slip discretized into $4 \times 6 = 24$ segments. The similarities between the connectivity matrices are apparent and especially between the 2D Burridge-Knopoff model and a strike-slip fault.}
\label{fig:gen_k_weights}
\end{figure*}

The matrix $\uubar{K}$ is called here \textit{connectivity matrix} and contains information on how the blocks are connected together. Various geometrical configurations, such as those presented in Fig.~\ref{fig:gen_slider}, can be described by adequately adjusting the components of the connectivity matrix. Fig.~\ref{fig:gen_k_weights} shows graphically the connectivity matrix for the 1D Burridge-Knopoff model, its 2D generalization in a $4\times6$ grid and a strike-slip fault discretized in $4\times6$ segments (blocks) as described in \citeA{Rice1993,Chinnery1963}. The similarities between the connectivity matrices are apparent and especially between the 2D Burridge-Knopoff model and a strike-slip fault (embedded in a semi-infinite elastic domain). All the blocks at the edges of the 2D Burridge-Knopoff model were considered to be connected with springs and dashpots with the ground, except the upper six which are connected only with the other blocks in order to qualitatively approximate the free boundary of the strike-slip fault discretization based on the \citeA{Rice1993,Chinnery1963} approach.

\section{Dynamic simulations and SOC}
\label{sec:simulations}
We consider that the driver plate is moving under constant velocity, which is several orders of magnitude lower than the velocities of the blocks that are developed during abrupt sliding. Consequently, we can assume that the driver plate remains still during the sliding events. This situation is inspired by the far-field earth's tectonic movement, which is several of orders of magnitudes lower (some centimeters per year) than the seismic slip velocities developed during earthquakes that can reach up to one meter per second. 

For the simulations, we first calculate the minimum displacement of the driver plate that can trigger the sliding of at least one block. In this way we avoid simulating the slow-dynamics \cite{Stefanou2019}, quasi-static behavior of the system and we only integrate numerically the dynamic equations of motion of the system for determining its fast dynamic, unstable response. After each dynamic event, the system reaches a new equilibrium (local minimum potential energy state). The slip of the blocks is recorded and a random small overshoot in their displacements is considered \cite<see also>[]{Brown1991,Rundle1991}. The random overshoot is not the same between the blocks and varies from zero to $20\%$ of its slip during the previous event. The random overshoot embodies several uncertainties of the system related to its elastic parameters, initial conditions and frictional properties, among others.

The frictional properties of the blocks can be uniform or randomly chosen from a distribution. Slip or slip-rate softening is required to render the system unstable and lead to SOC behavior. Here, we use slip weakening friction as in \citeA{Huang1992, Stefanou2019}. Simulations with slip-rate weakening would give similar results \cite<cf.>[]{Carlson1989,Huang1992a}. In particular, the friction coefficient evolves from its static value ($\mu_s$) to its kinetic one ($\mu_k$). In the simulations presented here $\mu_s=0.8$ and $\mu_k=0.5$. The friction drop occurs in a characteristic distance equal to $\hat D_c=.01$. The damping ratio $\zeta$ is set equal to one and $\frac{k_i^c}{k_{ij}^l}=2$.

Each slip event can involve sliding of a single block, a cluster of some blocks or of all the blocks of the system. After each slip event the friction coefficient of each block is restored to its static friction value and a new period of quiescence takes place as shown in Fig.~\ref{fig:events}.

The simulation procedure is summarized as follows:
\begin{enumerate}
  \item \textit{Quiescence period:} Determine $\ubar{\hat u}_\infty$ that renders the system unstable by solving $\ubar{\Psi}=\ubar{0}$ and set the driver plate displacement equal to $\displaystyle \hat u_\infty=\min_{\forall i}(\{\ubar{\hat u}_\infty\}_i)$
  \item \textit{Sudden slip event:} Integrate numerically the dynamic equations of motion \eqref{eqn:dynamics} to determine slip $\hat u_i$, until $\displaystyle \max_{\forall i}(\hat v_i) \le \textrm{threshold}$. The threshold was set equal to $0.02$, which is much smaller than the maximum velocity of blocks during unstable sliding (see Fig.~\ref{fig:events}b-d) .
  \item \textit{Healing:} Set block velocities equal to zero and update their positions $\hat u^0_i=\hat u_i+\tilde u_i$, where $\tilde u_i$ is a random overshoot as described above. Restore the friction of coefficient from $\mu_k$ to $\mu_s$.
\end{enumerate}
Repeat 1 to 4 and record events. A sequence of $N=10000$ events were simulated for calculating the frequency-size statistics presented in Fig.~\ref{fig:powerlaw}. The simulation of more events ($20000$) lead to almost identical results.

\section{Robust state feedback stabilization}
\label{sec:control_stabilization}
\begin{figure}
\centering
\includegraphics[width=.7\linewidth]{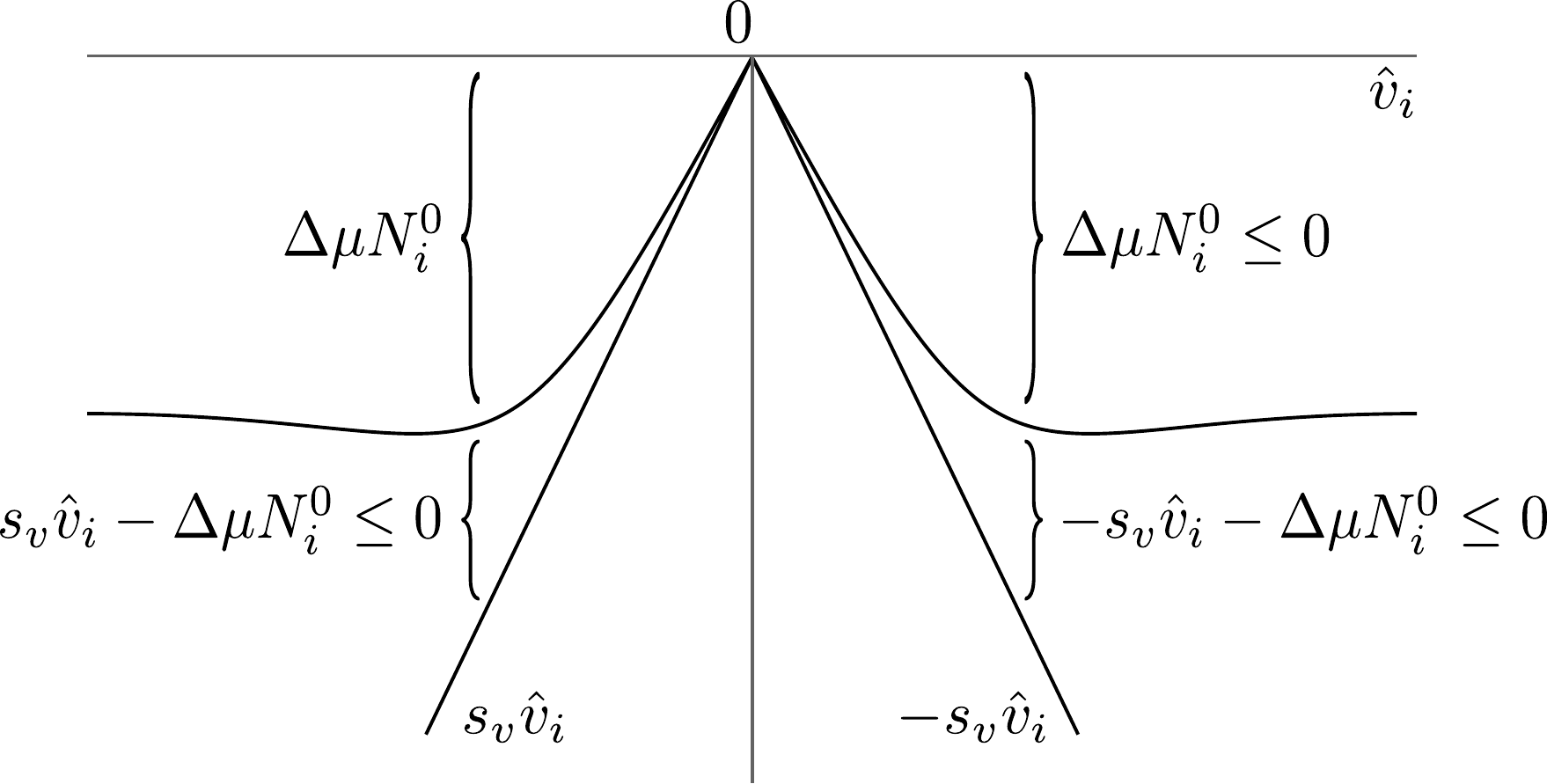}
\caption{Schematic representation of maximum slip-rate weakening slope $s_v \ge 0$ that bounds frictional weakening. The ordinate represents the various mathematical quantities presented in the appendix and shows schematically how they are bounded. The same bounds hold for slip weakening $s_u \ge 0$. Friction is always opposite to slip velocity.}
\label{fig:friction_limits}
\end{figure}

After some algebra  \eqref{eqn:dynamics} (or \eqref{eqn:dynamics_simple}) is written as follows:
\begin{equation}
\ubar{x}'= \uubar{A} \ubar{x} +\uubar{B} \ubar{\overline{P}}+\ubar{\overline{\Psi}},
\label{eqn:dynamics_control}
\end{equation}
where $\uubar{A}=\uubar{G}+\uubar{S}$, $\uubar{S}=\begin{bmatrix}
\uubar{O} & \uubar{O} \\
s_u\uubar{I} & s_v\uubar{I}
\end{bmatrix}$, $s_u$ and $s_v$ are respectively the absolute values of the minimum slip and slip-rate softening rates of the frictional law (see Fig.~\ref{fig:friction_limits}),
$\uubar{B}=\begin{bmatrix}
\uubar{O}\\
\mu_{min}\uubar{I}
\end{bmatrix}$, $\displaystyle \mu_{min}=\min_{\forall i, \forall u_i, \forall v_i}\mu_i > 0$ and $\ubar{\overline{\Psi}}=\ubar{\overline{\Psi}}_1+\ubar{\overline{\Psi}}_2$. \\ $\ubar{\overline{\Psi}}_1=\uubar{\widetilde{B}}\ubar{\overline{P}}$ with $\uubar{\widetilde{B}}=\begin{bmatrix}
\uubar{O}\\
\uubar{\tilde{\mu}}
\end{bmatrix}$, $\uubar{\tilde{\mu}}$ a diagonal matrix with diagonal elements such as $\{\uubar{\tilde{\mu}}\}_{ii}=\mu_i-\mu_{min}$ and $\ubar{\overline{\Psi}}_2=-\begin{bmatrix}
\uubar{O}\\
\overline{\uubar{s}}\left(\uubar{\Delta\mu}\ubar{N}^0+\overline{\uubar{s}} \uubar{S} \ubar{x} \right)
\end{bmatrix}$, where $\overline{\uubar{s}}$, $\uubar{\Delta\mu}$ are diagonal matrices with diagonal elements $\{\uubar{\overline{s}}\}_{ii}=\text{sgn}(\hat v_i)$, $\{\uubar{\Delta\mu\}}_{ii}=\mu_i-\mu_{si}$, respectively, $\ubar{N}^{0}$ a vector, with elements $\{\ubar{N}^{0}\}_{i}=N_i^0$, $N_i^0$ is a reference normal (effective) force applied at block $i$ and $\text{sgn}(.)$ is the sign function. Finally, $\ubar{\overline{P}}=\overline{s} \ubar{P}$ (equivalently the input $\ubar{P}=\uubar{\overline{s}}\ubar{\overline{P}}$), that needs to be determined for assuring asymptotic stability.

Let the scalar function $V(\ubar{x})=\frac{1}{2}\ubar{x}^T \uubar{\Theta}\ubar{x}>0$ for all non-zero $\ubar{x}\in\mathbb{X}\subset \mathbb{R}^{2n}$ ($n$ is the number of blocks) and $V(\ubar{0})=0$. Under these conditions, $\uubar{\Theta}$ is \textit{positive definite}, $\uubar{\Theta}\succ\uubar{0}$, \cite{Brauer1969} (or \textit{negative definite} if $\ubar{x}^T \uubar{\Theta}\ubar{x}<0$ $\forall\ubar{x}\in\mathbb{X}\subset \mathbb{R}^{2n}\backslash 0$, i.e. $\uubar{\Theta}\prec\uubar{0}$). Moreover let:
\begin{equation}
\overline{P}=-\frac{1}{2}\uubar{B}^T\uubar{\Theta} \left(\ubar{x}+\ubar{\tilde{x}} \right)
\label{eqn:control_law}
\end{equation}
be the control law. $\ubar{\tilde{x}}$ is a perturbation (see Fig.~\ref{fig:controller}). We search $\Theta$ such that the closed-loop system $\Sigma(P_{\ell},C)$ (Fig.~\ref{fig:controller}) can be \textit{asymptotically stable} at $\ubar{x}=\ubar{0}$. According to Lyapunov's stability theorem \cite<see \textit{Lyapunov's Second Method,}>[]{Brauer1969}, if there exists $V(\ubar{x})>0$ for which $V'(\ubar{x})$ is strictly negative $\forall \ubar{x} \in \mathbb{X}\subset\mathbb{R}^{2n}\backslash\ubar{0}$, then the origin of the system, $\ubar{x}=\ubar{0}$, is asymptotically stable. If $\mathbb{X}$ extends over the whole real $2n$-dimensional Euclidean space, then the origin is \textit{globally} asymptotically stable. Differentiating $V(\ubar{x})$ with respect to time, using \eqref{eqn:dynamics_control} and the symmetry of $\uubar{\Theta}$ we obtain:
\begin{equation}
V'(\ubar{x})=-\frac{1}{2}\ubar{x}^T \uubar{Q} \ubar{x}
-\frac{1}{2}\ubar{x}^T \uubar{\Xi} \ubar{\tilde{x}}
+\ubar{x}^T \uubar{\Theta} \ubar{\overline{\Psi}},
\label{eqn:control_Vdot}
\end{equation}
where $\uubar{\Xi}=\uubar{\Theta}\uubar{B}\uubar{B}^T\uubar{\Theta}$ and $\uubar{Q}$ satisfies the algebraic Riccati equation:
\begin{equation}
\uubar{A}^T\uubar{\Theta}+\uubar{\Theta}\uubar{A}-\uubar{\Theta}\uubar{B}\uubar{B}^T\uubar{\Theta}=-\uubar{Q}.
\label{eqn:control_Riccati}
\end{equation}
$\uubar{Q}$ is selected to be real, symmetric and positive definite ($\uubar{Q}\succ\uubar{0}$). In the numerical examples $\uubar{Q}$ was taken equal to the identity matrix but any other matrix could be selected as well provided that it satisfies the aforementioned conditions. Notice that in the frame of linear systems, which though is not our case, the matrix $\uubar{Q}$ can be selected optimally in order to satisfy some engineering criteria (see \textit{Linear Quadratic Control} problem). For a given $\uubar{Q}$ exists a unique positive definite and symmetric $\uubar{\Theta}$ satisfying the algebraic Riccati equation.

Therefore, a sufficient condition for the system to be asymptotically stable ($V'(\ubar{x})<0$) is the third and second terms of the right hand side of \eqref{eqn:control_Vdot}, to be negative or zero, i.e. $\quad \forall \ubar{x} \in \mathbb{X}\subset\mathbb{R}^{2n}\backslash\ubar{0}$:
\begin{equation}
\ubar{x}^T \uubar{\Theta} \ubar{\overline{\Psi}}=
\underbrace{\ubar{x}^T \uubar{\Theta} \ubar{\overline{\Psi}}_1}_{\Omega_1}
+\underbrace{\ubar{x}^T \uubar{\Theta} \ubar{\overline{\Psi}}_2}_{\Omega_2}
\le 0
\label{eqn:control_Omega}
\end{equation}
and
\begin{equation}
\underbrace{-\frac{1}{2}\ubar{x}^T \uubar{\Xi} \ubar{\tilde{x}}}_{\Omega_3}\le 0.
\label{eqn:control_OmegaPert}
\end{equation}

Using \eqref{eqn:control_law} we obtain:
\begin{equation}
\Omega_1=\ubar{x}^T \uubar{\Theta}\overline{\ubar{\Psi}}_1=\ubar{x}^T \uubar{\Theta} \widetilde{\uubar{B}} \ubar{\overline{P}}=-\frac{1}{2}\ubar{y}^T \widetilde{\uubar{B}} \uubar{B}^T \ubar{y},
\label{eqn:control_W1a}
\end{equation}
where we set $\ubar{y}=\uubar{\Theta}\ubar{x}$. $\ubar{y}=\ubar{0}$ if and only if $\ubar{x}=\ubar{0}$ due to the positive definiteness of $\uubar{\Theta}$. By simple inspection, the matrix $\widetilde{\uubar{B}} \uubar{B}^T$ is \textit{positive semidefinite}, because it is the product of diagonal matrices with positive or zero diagonal elements (see definitions of the diagonal matrices $\widetilde{\uubar{B}}$ and $\uubar{B}$). A matrix $\uubar{D}$ is called positive (negative) semidifinite $\uubar{D} \succeq \uubar{0}$ ($\uubar{D} \preceq \uubar{0}$) if and only if $\ubar{x}^T \uubar{D}\ubar{x}\ge0$ ($\ubar{x}^T \uubar{D}\ubar{x}\le0$), $\forall \ubar{x} \in \mathbb{X}\subset\mathbb{R}^{2n}\backslash\ubar{0}$. Therefore, $\Omega_1\le0$.

The eigenvectors of $\uubar{\Theta}$ form an orthonormal base that can be used to express the arbitrary vectors $\ubar{x} \in \mathbb{X}\subset\mathbb{R}^{2n}\backslash\ubar{0}$, as $\ubar{x}=\sum_{k=1}^{2n}\alpha_k \ubar{\omega}^{(k)}$, with $\omega^{(k)}$ being the eigenvector $k$ of $\uubar{\Theta}$ and $\alpha_k$ real coefficients. 
The eigenvalues, $\lambda^{(k)}$ of $\uubar{\Theta}$ are all strictly positive, due to its positive definiteness, i.e. $\lambda^{(k)}>0$. Therefore:
\begin{equation}
\Omega_2=\ubar{x}^T \uubar{\Theta}\overline{\ubar{\Psi}}_2=\sum_{k=1}^{2n}\lambda^{(k)}\ubar{\Psi}^T_2 \ubar{x}.
\label{eqn:control_W2a}
\end{equation}
After some algebra,
\begin{equation}
\ubar{\Psi}^T_2 \ubar{x}=\left(-\uubar{\Delta\mu}\ubar{N}^0-\overline{\uubar{s}} \uubar{S} \ubar{x} \right)^T|\ubar{v}|,
\label{eqn:control_W2b}
\end{equation}
where $\{|\ubar{v}|\}_i=|\hat v_i|$ and $|.|$ the absolute value. As shown in Fig.~\ref{fig:friction_limits}, $s_u\ge0$ and $s_v\ge0$ are such that each component of the vector $-\uubar{\Delta\mu}\ubar{N}^0-\overline{\uubar{s}} \uubar{S} \ubar{x}$ is negative. Consequently, $\Omega_2\le 0$ and \eqref{eqn:control_Omega} is satisfied. 

After some algebra \eqref{eqn:control_OmegaPert} takes the following form:
\begin{equation}
\left(\ubar{x}+\ubar{\tilde{x}}\right)^T\uubar{\Xi} \left(\ubar{x}+\ubar{\tilde{x}}\right)-\left(\ubar{x}-\ubar{\tilde{x}}\right)^T\uubar{\Xi} \left(\ubar{x}-\ubar{\tilde{x}}\right)\ge 0
\label{eqn:Omega3a}
\end{equation}
$\uubar{\Xi}$ is positive semidefinite due to the definition of matrix $\uubar{B}$ and has eigenvalues $\xi^{(i)}\ge 0$. Therefore, the sufficient condition $\Omega_3\le 0$, \eqref{eqn:control_OmegaPert}, becomes:
\begin{equation}
\sum_{k=1}^{2n}\xi^{(k)}x_k \tilde{x}_k \ge 0
\label{eqn:Omega3b}
\end{equation}
Herein, we consider small input perturbations satisfying the above inequality. Hence $\Omega_3\le 0$. In any other case, the above inequality defines a subset of the Euclidean space for $\ubar{x}$, in which asymptotically stability is guaranteed.

As a result, the closed-loop system has $V'(\ubar{x})<0$ and it is globally asymptotically stable under the control law of \eqref{eqn:control_law} with $\uubar{\Theta}$ satisfying \eqref{eqn:control_Riccati}. The input takes the final form:
\begin{equation}
\ubar{P}=-\frac{1}{2}\uubar{\overline{s}}\uubar{B}^T\uubar{\Theta} \ubar{x}_c,
\label{eqn:control_law2}
\end{equation}
where $\ubar{x}_c=\ubar{x}+\ubar{\tilde{x}}$.

\section{Robust tracking}
\label{sec:control_tracking}
Let $\ubar{r}_d=\ubar{r}_d(t)$ be a vector describing the desired displacements of the blocks. We want to minimize the error $\uubar{C} \ubar{x}-\ubar{r}_d$, where $\uubar{C}=\begin{bmatrix}
\uubar{I}  & \uubar{O}
\end{bmatrix}$.

Here we use the approach of \textit{integral action} \cite{Ka} and we augment the system with the equation:
\begin{equation}
\ubar{e}'=\uubar{C}\ubar{x}- \ubar{r}_d.
\label{eqn:control_integrator}
\end{equation}
Let also $\ubar{x}^{eq}$, $\ubar{e}^{eq}$ and $\ubar{P}^{eq}$ be such that:
\begin{equation}
\begin{split}
\ubar{0}&= \uubar{A} \ubar{x}^{eq} +\uubar{B} \ubar{\overline{P}}^{eq}+\ubar{\overline{\Psi}}^{eq} \\
\ubar{0}&=\uubar{C}\ubar{x}^{eq}- \ubar{r}_d.
\end{split}
\label{eqn:control_steadystate}
\end{equation}
$\ubar{x}^{eq}$, $\ubar{e}^{eq}$ and $\ubar{P}^{eq}$ exist for the physical system at hand.

Applying the transformation $\ubar{z}=\ubar{x}-\ubar{x}^{eq}$, $\ubar{\xi}=\ubar{e}-\ubar{e}^{eq}$, Eqs. (\ref{eqn:dynamics_control},~\ref{eqn:control_integrator}) become:
\begin{equation}
\ubar{w}'= \uubar{A}_a \ubar{w} +\uubar{B}_a \Delta \ubar{\overline{P}}+\ubar{\overline{\Psi}}_a,
\label{eqn:dynamics_control_tracking}
\end{equation}
where $\ubar{w}=\begin{bmatrix}
\ubar{z}\\
\ubar{\xi}
\end{bmatrix}$,
$\uubar{A}_a=\begin{bmatrix}
 \uubar{A} & \uubar{O}\\
 \uubar{C} & \uubar{O}
\end{bmatrix}$, 
$\uubar{B}_a=\begin{bmatrix}
 \uubar{B} \\
 \uubar{O}
\end{bmatrix}$, $\ubar{\Psi}_a=\begin{bmatrix}
 \ubar{\Psi}-\ubar{\Psi}^{eq} \\
 \ubar{O}
\end{bmatrix}$ and $\Delta \ubar{P}=\ubar{P}-\ubar{P}^{eq}$.

The above system has the same form with \eqref{eqn:dynamics_control} and the same analysis as above can be carried out for determining the controller that assures its stabilization, leading to:
\begin{equation}
\ubar{\Delta P}=-\frac{1}{2}\uubar{\overline{s}}\uubar{B}^T_a\uubar{\Theta}_a \ubar{w}
\label{eqn:control_law3}
\end{equation}
or, equivalently, 
\begin{equation}
\ubar{P}=-\frac{1}{2}\uubar{\overline{s}}\uubar{B}^T_a\uubar{\Theta}_a \ubar{x}_a,
\label{eqn:control_law4}
\end{equation}
where $\ubar{x}_a=\begin{bmatrix}
\ubar{x}\\
\ubar{e}
\end{bmatrix}$ and $\uubar\Theta_a$ is the solution of the Riccati equation. \eqref{eqn:control_Riccati}, but for $\uubar{A}_a$, $\uubar{B}_a$ and $\uubar{Q}_a$ instead of $\uubar{A}$, $\uubar{B}$ and $\uubar{Q}$, respectively. Similarly, $\uubar{Q}_a$ is selected to be real, symmetric and positive definite ($\uubar{Q}_a\succ\uubar{0}$).

By regulating the input pressure as in \eqref{eqn:control_law4}, the system can be driven to the desired position ensuring robust tracking. Indeed, due to asymptotic stability of \eqref{eqn:dynamics_control_tracking} under \eqref{eqn:control_law3} or \eqref{eqn:control_law4}, $\displaystyle \lim_{t \rightarrow \infty} \ubar{\xi}'=\lim_{t \rightarrow \infty} \ubar{e}'=\ubar{0}$ and, therefore, $\displaystyle \lim_{t \rightarrow \infty} \uubar{C}\ubar{x}=\ubar{r}_d$. 

\section{Numerical implementation}
The numerical integration of the ordinary differential equations presented herein was performed using \textit{SciPy} \cite{Virtanen2020} and the \textit{LSDOA} implicit algorithm \cite{Hindmarsh1983,Petzold1983}. The designed controller was programmed in \textit{Python 3} \cite{VanRossum1995} and the algebraic Riccati equation was solved using the \textit{Python Control Systems Library} \cite{PythonControl_0.8.3}. The programs are available in \textit{Jupyter notebooks} \cite{Kluyver:2016aa} upon request.

\bibliography{Papers-JGR}

%
%
%
%
%

\end{document}